\documentclass[aps,pra,twocolumn,longbibliography, superscriptaddress]{revtex4-1}%
\usepackage{graphics}
\usepackage{amsmath}
\usepackage{graphicx}
\usepackage{amsfonts}
\usepackage{amssymb}
\usepackage{float}
\usepackage{longtable}
\usepackage{epsfig}
\usepackage{color}
\usepackage{latexsym}
\usepackage{theorem}
\usepackage{bbm}
\usepackage{bm}
\usepackage{verbatim}
\usepackage{times}
\usepackage{psfrag}%
\usepackage[dvipsnames]{xcolor}
\graphicspath{ {./figs/} }
\newcommand{\je}[1]{{\color{black} #1}}

\newcommand{\eqn}[1]{\begin{eqnarray} \newline #1 \end{eqnarray}}
\newcommand{\bk}[1]{\left ( #1 \right )}

\usepackage{hyperref}
\hypersetup{
    colorlinks=true,
    linkcolor=blue,
    filecolor=magenta,      
    urlcolor=cyan,
}

\DeclareMathOperator*{\Tr}{Tr}
\newtheorem{theorem}{Theorem}
\newtheorem{definition}{Definition}
\begin{document}
\title{Rate limits in quantum networks with lossy repeaters}
\author{Riccardo Laurenza}
\affiliation{Dahlem Center for Complex Quantum Systems, Freie Universit{\"a}t Berlin, 14195 Berlin, Germany}
\author{Nathan Walk}
\affiliation{Dahlem Center for Complex Quantum Systems, Freie Universit{\"a}t Berlin, 14195 Berlin, Germany}
\affiliation{Toshiba Research Europe Limited, 208 Cambridge Science Park, Cambridge CB4 0GZ, United Kingdom}
\author{Jens Eisert}
\affiliation{Dahlem Center for Complex Quantum Systems, Freie Universit{\"a}t Berlin, 14195 Berlin, Germany}
\affiliation{Helmholtz-Zentrum Berlin f{\"u}r Materialien und Energie, Hahn-Meitner-Platz 1, 14109
Berlin, Germany}
\affiliation{Fraunhofer Heinrich Hertz Institute, 10587 Berlin, Germany}
\author{Stefano Pirandola}
\affiliation{Department of Computer Science, University of York, York YO10 5GH, United Kingdom}

\begin{abstract}
The derivation of ultimate limits to communication over certain quantum repeater networks have provided extremely valuable benchmarks for assessing near-term quantum communication protocols. However, these bounds are usually derived in the limit of ideal devices and leave questions about the performance of practical implementations unanswered. To address this challenge, we quantify how the presence of loss in repeater stations affect the maximum attainable rates for quantum communication over linear repeater chains and more complex quantum networks. Extending the framework of node splitting, we model the loss introduced at the repeater stations and then prove the corresponding limits. In the linear chain scenario we show that, by increasing the number of repeater stations, the maximum rate cannot overcome a quantity which solely depends on the loss of a single station. We introduce a way of adapting the standard machinery for obtaining bounds to this realistic scenario. The difference is that whilst ultimate limits for any strategy can be derived given a fixed channel, when the repeaters introduce additional decoherence, then the effective overall channel is itself a function of the chosen repeater strategy (e.g., one-way versus two-way classical communication). Classes of repeater strategies can be analysed using additional modelling and the subsequent bounds can be interpreted as the optimal rate within that class. 
\end{abstract}
\maketitle
\section{Introduction}
Quantum communication is one of the most practically relevant applications of the quantum technologies, offering the perspective of secure communication
based on physical laws
\cite{Pirandola:20,RevModPhys.92.025002,RevModPhys.74.145,TeleportationReview,Renner2021,Principles,Werner2001,RevSpace,MohsenBook}. While security can be proven to hold 
under enormously generous and general 
conditions, it can only be guaranteed for 
sufficiently low levels of loss. For short distances, this does not constitute a technological challenge.
%
%
However, for large distances, secure quantum communication becomes very challenging, since all loss has to be attributed to an eavesdropper and this prevents achieving arbitrarily high rates of secure bits. Similar limitations arise for the maximum  attainable rates for quantum information (qubits) transmission entanglement (ebits) distribution over lossy bosonic channels that conveniently describe optical fibres or free-space links.
%
More specifically, it has been established that, for any point-to-point transmission protocol over a lossy bosonic channel with transmissivity equal to 
$\eta\in (0,1)$, allowing the two parties to exploit unlimited two-way classical communications, the maximum achievable rates for key generation, entanglement distribution, and quantum bit transmissions, are all equal to the \emph{repeaterless PLOB bound} $-\log_2(1-\eta)$~\cite{PLOB}. 

This severe limitation of direct point-to-point transmission is not a road block, however. Intermediate stations, referred to as \emph{quantum repeaters} \cite{RevModPhys.83.33}, 
can overcome this limitation and, in principle, allow communication over arbitrary
distances. Since the appearance of the first quantum repeater proposal~\cite{Briegel}, the goal of extending the distance at which two parties can faithfully share a secret key or entanglement has stimulated a plethora of repeater-assisted quantum communication schemes. The difficulty of assessing precise rates of (quantum) information 
transmission and specifically
of key rates gives rise
to the necessity to identify
bounds that are 
agnostic
to the specific implementation
chosen. Only in such a realm,
can ultimate bounds for quantum
communication be established.
Along this line of thought, a fundamental result about the rates at which two end-nodes in a linear repeater chain can transmit quantum information, distribute entanglement, or generate a secret key has been established in Ref.~\cite{PirNetwork}. In particular, when two users, say Alice and Bob, are connected by a line of $N-1$ middle repeater nodes, linked together through $N$ optical lossy fibres, the 
quantum/private capacity
of the linear repeater chain, i.e., the ultimate rate for repeater-assisted quantum or private communication
between the two end-users is given by~\cite[Eq.~(9)]{PirNetwork}
\begin{equation}\label{idealRep}
C(\eta,N)=-\log_2(1-\sqrt[N]{\eta})~,
\end{equation}
where $\eta>0$ is the total Alice-to-Bob fibre transmissivity. This expression is derived by exploiting a combination of tools that we briefly recall in the appendices.

From the conceptual point of view, a traditional quantum repeater is a scheme in which entanglement is first distributed to intermediate nodes. Then, the quality is improved by means of entanglement distillation, transforming a collection of weakly entangled states into a smaller number of more highly entangled states. In the final step, one performs sequential entanglement swapping, bringing quantum systems together that have no joint past, to entangle the anticipated nodes. More recently, repeater schemes based on quantum error correction have appeared, where the repeater nodes instead contain quantum gates to implement the necessary operations. Nevertheless, in the case of ideal devices, both varieties have a performance that is upper bounded by Eq.~(\ref{idealRep}).

It is important to stress that, for a fixed total transmissivity $\eta$ but large number of repeaters, the end-to-end capacity $C(\eta,N)$ 
diverges as $\log_2N$. 
This feature is immediately connected to the fact that the repeaters are assumed to be ideal, i.e., lossless. Under a more realistic point of view each repeater in the linear 
\je{repeater chain}
must be characterised by imperfect components which introduce noise and decoherence to the stored and transmitted quantum states. For instance these internal losses could be the effect of non-unit detection efficiencies, channel-memory coupling losses, memory loading and readout efficiencies. Furthermore, detrimental effects can be introduced by the quantum memories at the nodes while the quantum systems are stored before the on-demand retrieval.

Here, we explicitly account for this crucial aspect and we consequently derive the end-to-end repeater capacity of a lossy bosonic quantum network where the repeater stations are also affected by internal loss. Although loss is not the only source of decoherence, it is often the dominant factor and is an excellent first approximation for an optical fibre channel. Furthermore, using lossy channels as a model for imperfections within the repeater nodes will facilitate derivation of exact formulae for the various capacities of interest. 

Our derivation is carried out for the various types of 
\emph{routing} (single- and multi-path). Our bound hence captures rather general classes of repeater schemes and can be seen as an analog
of the repeater-less PLOB in the presence of lossy repeater stations. Given a fixed amount of loss for each repeater node we can immediately evaluate our bounds as a function of transmission losses. However, in real implementations the loss in each node is itself typically a function of the transmission losses and the type of repeater strategy employed. The paradigmatic example is the loss induced by a quantum memory where the necessary storage time usually increases with transmission loss. In this scenario, the two major classes of strategy are those that require either one-way or two-way classical communication, where the latter corresponds to a longer typical waiting and hence a lossier effective channel.

To exemplify these findings, we
show that, considering a realistic time-dependent model of decoherence for a single repeater station, the achievable rates not only beat the benchmark of the repeater-less PLOB bound, but they are also not that far from the upper limit provided by our revised lossy-repeater capacity. As an example we consider polarisation-based BB84 key distribution protocol over a single repeater node using a simple entanglement swapping protocol with Rubidium memories and show it scales as one quarter of the optimal possible rate for such schemes.
Finally, recent years have enjoyed considerable
interest in identifying practical schemes
for routing in multi-partite 
\emph{quantum networks}
\cite{PirNetwork,QuantumInternet,PhysRevA.93.032302,EppingA,HahnPappaEisert,NetSquid,PhysRevA.103.032610}. Our results
are general enough to accommodate such situations, as we show.

\section{Node splitting}


Let us consider a linear sequence $\{s_0,\ldots,s_{N}\}$ of $N-1$ repeater nodes, where Alice and Bob, the two end stations, are identified with $s_0$ and $s_N$, respectively. 
We assume that each station $s_i$ in the chain is connected to $s_{i+1}$ through an optical fibre described by a Gaussian lossy channel 
\cite{GaussianChannel} $\mathcal L_i$ with transmissivity $\eta_i$, for $i=0,\dots, N$. Thus, the total transmissivity of the link (e.g., an optical fibre
providing the communication channel) connecting Alice and Bob is  $\eta=\prod_i\eta_i$. 
%
%
%
Each node $s_i$ has internal losses that can be quantified by a global transmissivity $\tau_i\in[0,1]$ 
obtained by the product of single inefficiencies. 
In this way we can describe the effect of the node on the incoming quantum systems as another Gaussian lossy channel, mathematically described as a beam splitter mixing the input system with an environment in the vacuum
\begin{equation}\label{BS}
\hat{x}_{\text{out}}=\sqrt\tau_i\hat{x}_{\text{in}}+\sqrt{1-\tau_i}\hat{x}_{\text{vac}}~.
\end{equation}
We can further distinguish two different contributions in $\tau_i$: a {\itshape transmitting} efficiency $\tau_i^t$, and a {\itshape receiving} efficiency $\tau_i^r$. The former is associated for instance with the overall effects of a source efficiency (e.g. photon creation efficiency), a memory read-out efficiency and a memory-channel interface efficiency. The latter involves a detector efficiency, a memory write-in efficiency and channel-memory coupling efficiency in some fashion.

\begin{figure}[htbp]
\vspace{0.1cm}
\par
\begin{center}
\includegraphics[width=0.49 \textwidth]{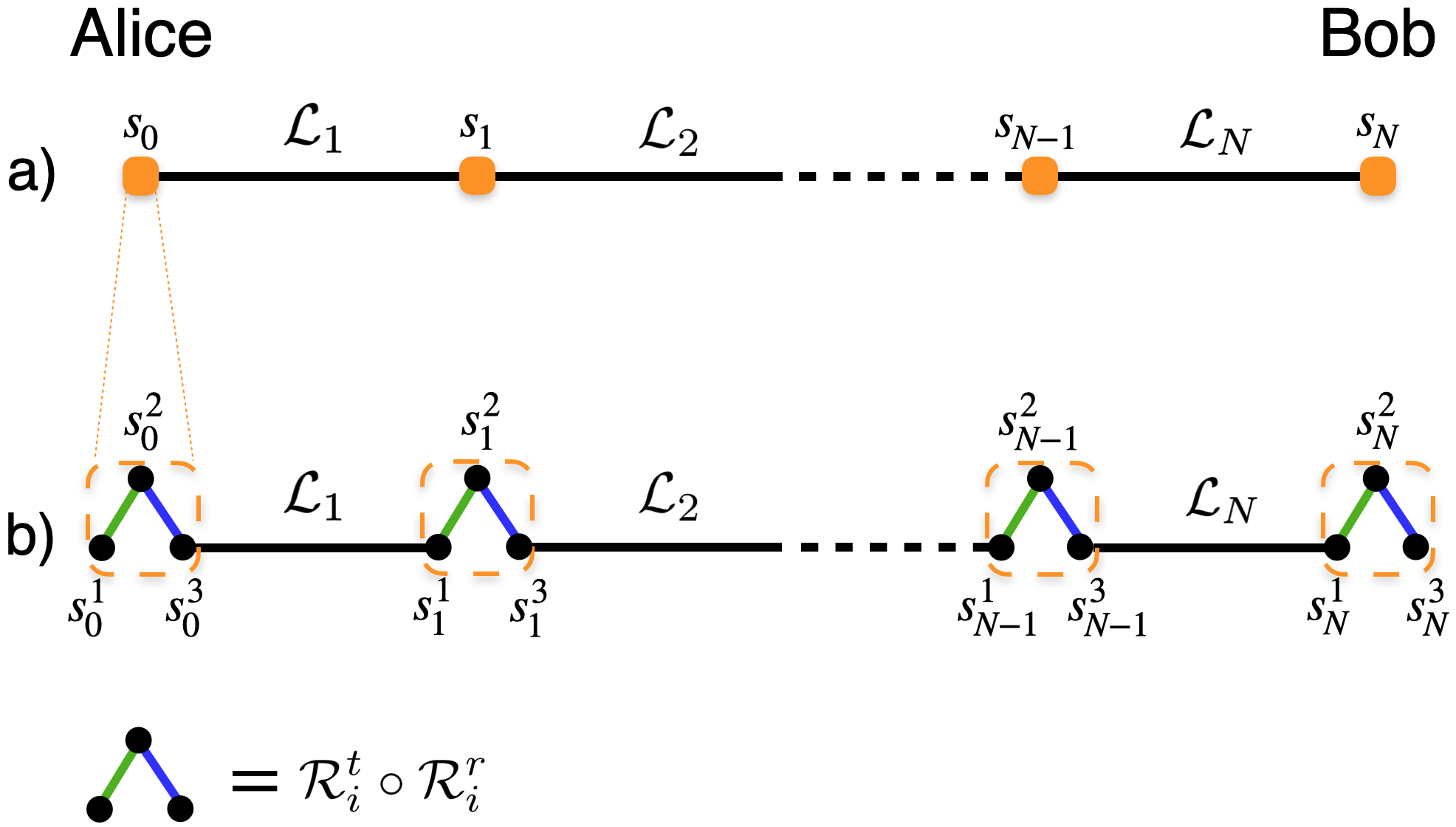} \vspace{-0.6cm}
\end{center}
\caption{Node splitting in a repeater chain. a) $N-1$ repeater stations $s_i$ are linked together to form a linear network (chain) between $s_0$ (Alice) and $s_N$ (Bob). The end-to-end transmissivity is $\eta=\eta_1\eta_2\cdots\eta_N$, where $\eta_i>0$ is the transmissivity of the single link described by the quantum lossy channel $\mathcal L_i$.  b) Node splitting of the linear network. Each node $s_i$ is split into three children nodes $\{s_i^1, s_i^2, s_i^3\}$ and two links $s_i^1-s_i^{2}$, $s_i^2-s_i^{3}$ are added. The overall effect of the internal losses in the $i$-th node is then described by the composition $\mathcal R_i^t\circ\mathcal R_i^r$ of two additional quantum lossy channels.} 
\label{Fig:lossyrep}
\vspace{0.3cm}
\end{figure}

To account for the internal lossy features in the various stations, we perform the \emph{node splitting} depicted in Fig.~\ref{Fig:lossyrep}. 
We split the generic node $s_i$ into three ``children'' nodes $s_i^k$ ($k=1,2,3$), which are then linked together through a composition of two lossy channels $\mathcal R_{s_i^2\rightarrow s_i^3}^t$ and $\mathcal R_{s_i^1\rightarrow s_i^2}^r$, with associated transmissivities $\tau_i^{t,r}$. 
Combining these internal channels with $\mathcal L_i$ associated to the $i$th link, we can model the linear network with noisy quantum repeaters as a sequence of composite quantum channels. More precisely, we can identify a building-block channel, so that the linear network can be described as the collection $\{\mathfrak C_i\}_i$ of the following composite quantum channels  (see Fig.~\ref{Fig:lossyrep})
\begin{equation}\label{DefChannel}
\mathfrak C_i=\mathcal R_{s_{i+1}^1\rightarrow s_{i+1}^2}^r\circ\mathcal L_{i+1}\circ\mathcal R_{s_{i}^2\rightarrow s_{i}^3}^t,
\end{equation}
for $i=1,\ldots,N-1$, while for the two end-nodes we set $\mathcal R_{s_0^1\rightarrow s_0^2}^r=\mathcal R_{s_N^2\rightarrow s_N^3}^t=\mathcal I$, where $\mathcal I$ is the identity channel. To simplify notation, we rename $\mathcal R_{s_i^k\rightarrow s_i^{k+1}}^{r,t}=\mathcal R_{i}^{r,t}$.

By means of the decomposition in Eq.~(\ref{DefChannel}), we are able to apply the machinery developed in Ref.~\cite{PirNetwork} to our scenario so we can derive a single-letter upper bound on the secret-key capacity (and therefore on the two-way quantum capacity) of the lossy-repeater linear chain.
By performing an entanglement cut labelled by $i$, we disconnect the chain along the channel $\mathcal L_i$. In doing so, we create a bipartition $(A,B)$ of the chain, with  $A=\{s_0^1, s_0^2, s_0^3\ldots,s_i^1, s_i^2, s_i^3\}$ and $B=\{s_{i+1}^1, s_{i+1}^2, s_{i+1}^3\ldots,s_N^1, s_N^2, s_N^3\}$. This leads us
to formulate the following. 

\begin{definition}[Lossy repeaters]
The state shared by Alice and Bob at the output of the most general adaptive protocol over $n$ uses of the repeater chain 
is given by
\begin{equation}
\rho_{a, b}^n=\Lambda_i\left(\sigma_{\mathfrak C_i}^{\otimes n}\right),
\label{stretch}
\end{equation} 
where $\Lambda_i$ is a trace-preserving 
\emph{local operation with classical communication} 
(LOCC) while $\sigma_{\mathfrak C_i}$ is the Choi 
matrix of the channel $\mathfrak C_i$, which is defined as $\sigma_{\mathfrak C_i}:=(\mathcal I\otimes\mathfrak C_i)
(\Phi)$.\end{definition}

Here, $\mathcal I$ is the identity channel and $\Phi$ is a maximally-entangled state. More precisely, the above equation has to be intended as asymptotic, since for CV systems, the maximally entangled state is asymptotic and as a consequence the Choi matrix $\sigma_{\mathfrak C_i}$ is obtained as a limit. 
In the appendix, 
we give details on this argument.
We notice that for $i=0$ and for $i\geq1$, the quantum channel $\mathfrak C_i$ is a pure loss channel or a composition of two pure-loss channels respectively. Thus we can conclude that for any $i\geq0$, $\mathfrak C_i$ is a distillable channel, for which the two-way quantum and private capacities are identical and exactly determined, i.e.~\cite{PLOB}
\begin{equation}
\mathcal C(\mathfrak C_i)=E_R(\sigma_{\mathfrak C_i})=D_1(\sigma_{\mathfrak C_i})~,
\end{equation}
where $E_R(\sigma_{\mathfrak C_i})$ is the 
\emph{relative entropy of entanglement} of the Choi matrix $\sigma_{\mathfrak C_i}$ and $D_1(\sigma_{\mathfrak C_i})$ is the entanglement that can be \emph{distilled} from the Choi matrix with one-way, forward or backward, classical communication (see the appendix for a recap about these types of capacities).

By exploiting Theorem 7 in 
Ref.~\cite{PirNetwork}, we conclude that the two-way quantum/private capacity of the linear chain with lossy repeaters satisfies
\begin{equation}
\mathcal C(\{\mathfrak C_i\})=\min_{0\leq i\leq N}E_R(\sigma_{\mathfrak C_i})=\min_{0\leq i\leq N}\mathcal C(\mathfrak C_i).
\end{equation}
Using the PLOB bound and the fact that the transmissivity of a composition of  lossy channels is given by the product of the individual transmissivities, we can state the following theorem which generalizes the formula for ideal repeaters given in Ref.~\cite{PirNetwork},
\begin{theorem}[Lossy-repeater capacity]
The ultimate achievable  rate for repeater-assisted quantum/private communication between the two-end users of a linear network with $N-1$ lossy quantum repeaters connected by $N$ pure-loss channels 
is given by
\begin{equation}\label{chainCap1} 
\mathcal C(\{\mathfrak C_i\})=\min_i[-\log_2(1-\tau_i^t\tau_{i+1}^r\eta_{i+1})]~,
\end{equation}
i.e., it equals the minimum capacity of the channel $\mathfrak{C_i}$ describing the loss of node $i$, the pure loss channel $i+1$ and the loss of node $i+1$~\cite{note2}.
\end{theorem}
Let us assume that the end-users, Alice and Bob, are at a distance $L$ apart and connected by an optical fibre whose transmissivity $\eta$ decays exponentially as $\eta=e^{-\alpha L}$ (typically, $\alpha=0.2$dB/km). If $N-1$ lossy repeaters are inserted along the line, the optimal configuration is represented by equally spaced nodes at a distance $L_0=L/N$, so we have $\eta_{i}=\sqrt[N]\eta$ for each $i$. We can thus recast Eq.~(\ref{chainCap1}) as follows
\begin{equation}\label{chainCap2} 
\mathcal C(\{\mathfrak C_i\})=-\log_2(1-\widetilde\tau\sqrt[N]\eta)]~,
\end{equation}
where we have defined $\widetilde\tau:=\min_{i\geq0}\tau_i^r\tau_{i+1}^t$. 
For simplicitly, assume that all the nodes are built and equipped with the same components, i.e., $\tau_i^r=\tau^r$ and $\tau_i^t=\tau^t$~, for all $i\in[0,N]$. We then get
\begin{equation}
\mathcal C(\{\mathfrak C_i\})\rightarrow C_\tau(\eta,N)=-\log_2(1-\tau\sqrt[N]{\eta})~,
\end{equation}
where $\tau:=\tau^r\tau^t$.
If we now consider a large number of nodes we obtain the following expansion
\begin{equation}
C_\tau(\eta,N\gg1)\simeq-\log_2(1-\tau)+\frac{\tau\log_2\eta }{(1-\tau)N}~.
\end{equation}
We can thus see that, by increasing the number of repeaters between Alice and Bob, i.e., by taking the limit of $N\rightarrow\infty$, the lossy-repeater capacity is bounded by the quantity $-\log_2(1-\tau)$ that depends solely on the loss within a node. In other words, even if we are allowed to arbitrarily increase the number of repeaters on the line, the optimal rate will be anyway bounded by the inevitable internal loss which act as ultimate bottleneck in the process.

\section{Time-dependence and realistic repeater protocols on a quantum linear network}

While the above results illuminate the performance of repeater networks with imperfect devices, there is a certain tension between our desire to quantify the ultimate limits to communication whilst also providing formulae that are as relevant as possible to near term demonstrations. The reason for this is that the bounds derived above, whilst totally general in the sense of applying to an optimal two-way LOCC encoding and decoding strategies, only hold for a given channel. However, the effective channel induced by the decoherence of realistic repeater nodes is itself, to some extent, determined by the choice of repeater protocol. For example, the effective loss experienced by a system stored in a quantum memory is a function of the ratio between memory coherence time and the required storage time, but this latter quantity can change depending upon the chosen protocol. In this section, we address this issue for memory-based repeater protocols by taking into account the role played by time. Incorporating these effects is crucial to obtain tighter bounds that provide more accurate benchmarks for realistic repeater protocols with imperfect devices. This is also a powerful example of how our relatively simple model can be used to meaningfully compare different protocols, as the major differences between them often boil down to variations in timing.

Ultimately, some of the operations involved in the design of repeater-assisted quantum communication and entanglement distribution protocols are intrinsically probabilistic. In memory-based quantum repeater protocols, such fundamental operations are represented by heralded entanglement generation (and possibly distillation) between neighbouring nodes and swapping that transfers such entanglement to nodes at increasing distance. Thus, besides the time required for the transmission of the quantum information carriers and classical heralding signals, which is limited by the speed of light, a finite time is also needed while waiting for success of various operations at the different repeater stations. 


As a good first order approximation we can model the memories as time-dependent lossy channels with transmission given by (see, e.g., Ref.~\cite{Mem}),
\begin{equation}
\tau_{\mathrm{mem}}(t)=\tau_0e^{-t/t_c}~. \label{memloss}
\end{equation}
where $\tau_0$ is the maximum memory efficiency and $t_c$ is the coherence time.

The key task remaining to evaluate these bounds is to correctly model the expected storage time. Fortunately, this problem has been well studied in the literature \cite{Bernardes:2011ij, Shchukin:2019gs}. The situation can be analysed abstractly by defining the success probability of operations on one half of the repeater, $p$. The expected waiting time will be of the form $MT_0$, where $T_0>0$ is the time taken for one attempt and $M$ is the expected number of attempts. As a first illustration, consider the simplest, canonical setup of a linear chain with one repeater station and two segments separated by a total distance, $d$.

The quantities $T_0$ and $p$ are both influenced by the choice of repeater protocol. The minimal time unit, $T_0$, depends upon whether the central station is operating in a \emph{node-receives-photons} (NRP) or a \emph{node-sends-photons} (NSP) configuration  \cite{vanLoock:2020gt}. In the former case, $T_0$ is simply set by $R$, the clock speed of either the source or the local processing (e.g. memory write-in time), whichever is slower. Thus, $T_0^{\mathrm{NRP}} = 1/R$. In the NSP case, for sufficiently large distances, $T_0$ will instead be limited by the time taken to transmit quantum states from the central node to the end stations and subsequently receive a classical signal heralding their successful arrival and initiating the swap. This corresponds to the time to transmit twice over one segment such that $T_0^{\mathrm{NSP}} = \max \{1/R, d/c \}$. A final subtlety is that in the NSP configuration, even if the first attempt is successful, a state must still be stored at the central for the time taken for at least one quantum transmission and one classical signal heralding success, i.e., a total of $M +2$ time steps.  

The probabilistic elements that go into determining $p$ depend upon whether we think of a \emph{continuous variable} (CV) or \emph{discrete variable} (DV) scheme utilising single photon detection. In a DV scheme entanglement distillation can be avoided if desired and all that is strictly necessary is to store single photon until another has arrived that can be used to swap entanglement. Indeed it is this strategy that is currently pursued in state-of-the-art experiments \cite{Bhaskar:2020gh,Pu:2021tr} In this scenario, the probabilistic element is then simply the detection probability of a photon across a single link in the repeater chain and
\eqn{p = \sqrt{\eta}\tau^{t,\mathrm{eff}} \label{psimp}}
for a transmission node and an analogous expression for a receiving node. Here $\tau^{t,\mathrm{eff}}$ represents the efficiency of all of the elements in the transmitting nodes except the memory. These quantities, such as write-in, read-out or detection efficiencies, will all be time independent and can be captured by single constant. Note that certain nodes in a chain may not have memories.

In the CV case, the arrival of quantum information is deterministic, and the probabilistic element is the entanglement distillation operation. Once distillation has been successfully carried out on either input, that mode is stored until the mode on the other side as also been distilled and then entanglement is swapped. Again, whilst some distillation is essential, the exact amount is a free parameter leading to a trade-off between the success probability and amount of entanglement in the final state. There are only a relatively small number of CV repeater protocols \cite{Campbell:2012bq,Dias:2017jk,Furrer:2018im,Dias:2019tz} with arguably the most mature being those based upon a so-called 
\emph{noiseless linear amplifier} (NLA) \cite{proc-disc-2009,Xiang:2010ua}. The NLA acts with a gain $g$, and the success probability can be upper bounded by $p\leq 1/g^2$, although this bound can be very loose in some circumstances \cite{Pandey:2013wb}. This is in principle a free parameter, but a reasonable strategy would be to adjust the gain to reverse the effects of the expected losses prior to distillation. To undo a lossy channel of transmission $\tau$ requires a gain of $g^2 = 1/\tau$. For these choices, a CV distillation would have success probability upper bounded by (\ref{psimp}), exactly as for a DV scheme.

\begin{figure}[htbp]
\vspace{0.5cm}
\par
\begin{center}
\includegraphics[width=0.5 \textwidth]{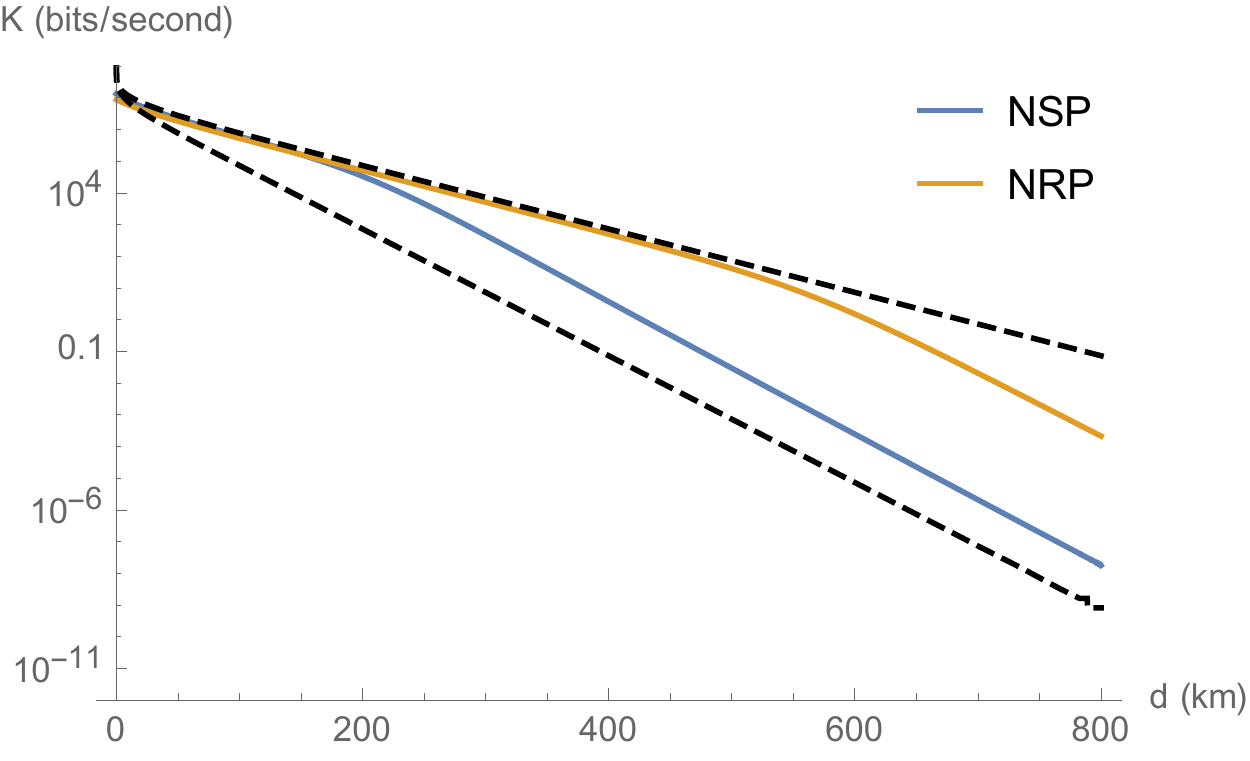} \vspace{-0.6cm}
\end{center}
\caption{Upper bounds to the secret key rate for both NSP and NRP protocols using Rubidium memories taken from Ref.~\cite{vanLoock:2020gt}. Parameters are: total efficiencies (which is what Ref.~\cite{vanLoock:2020gt} refers to as $P_{\rm link}$) of $(\tau^{\mathrm{eff}})^2 = 0.7$, a coherence time of 100 milliseconds and clock speed of $R = 5\times10^6$. The lower and the upper dashed black lines are respectively the repeaterless PLOB bound~\cite{PLOB} and the one-station repeater-assisted capacity~\cite{PirNetwork}.}
\label{Fig:rubidium}
\end{figure}

Putting all of this together, 
we compute the expected value of the memory transmission for the NRP and NSP configurations as \cite{Bernardes:2011ij, Shchukin:2019gs},
\eqn{\bar{\tau}_{\mathrm{mem}}^{\mathrm{NRP}} &=& \mathbb{E}\bk{\tau_0 e^{MT_0/t_c}}, \nonumber \\
&=& \frac{p}{2-p}\bk{\frac{2}{1 - e^{-T_0^{\mathrm{NRP}}/t_c}(1-p)  }- 1} \label{tmem} ,\\
\bar{\tau}_{\mathrm{mem}}^{\mathrm{NSP}} &=& \mathbb{E}\bk{\tau_0 e^{(M+2)T_0/t_c}} \nonumber \\
&=& \frac{p}{2-p}\bk{\frac{e^{-\frac{2 T_0^\mathrm{NSP}}{t_c}} \left((1-p)+e^{T_0^\mathrm{NSP}/t_c}\right)}{e^{T_0^\mathrm{NSP}/t_c}-(1-p)} } .
\nonumber
}
In either the NRP or NSP protocol, the total loss over one link will include whatever constant detection or coupling efficiencies are present along with the additional lossy channel induced by the memory, which will be at either the receiving or transmitting nodes. This means in either case we could write the total repeater losses as $\tau_i^t\tau_{i+1}^r = \tau_i^{t,\mathrm{eff}}\tau_{i+1}^{r,\mathrm{eff}}\tau_{\mathrm{mem}}$. Thus we can substitute (\ref{tmem}) into (\ref{chainCap1}) and, using parameters from Ref.~\cite{vanLoock:2020gt}, evaluate the bounds for some representative repeater platforms. We present the results for a platform based on Rubidium memories in Fig.~\ref{Fig:rubidium}. Note that because we are explicitly considering time in our analysis we are able to calculate rates in terms of bits per second, which is the quantity that is ultimately important for applications, as opposed to the more common bits per channel use.

Crucially, we see that our upper bound now has the same qualitative shape as a real repeater implementation. For short distances, where the storage times are small relative to the memory coherence time, the key rate scales as an ideal repeater with an offset due to extra losses at the station. However, for larger distances, the necessary storage time becomes comparable to the memory coherence time and thus the effective loss falls off exponentially faster. In this situation, the protocol fails to follow the ideal repeater scaling, regressing to scale similarly to the repeaterless bound. For certain system parameters our upper bounds can even drop below the repeaterless scaling as the waiting times for the NSP protocol cause additional losses that destroy any benefit for a repeater station.

Finally, we can also use our bounds to benchmark specific protocols carried out with the same system parameters. In Fig.~\ref{Fig:BB84comp}, we plot the ratio of a BB84 key rate using an entanglement swapping repeater protocol (see Appendix) to our lossy-repeater capacity given in Eq.~(\ref{chainCap1}). From this we can conclude that, over lossy repeater networks, standard BB84 and an entanglement swapping repeater is quite close to the optimal protocol, scaling identically for large distances and achieving slightly worse than one quarter of the optimal key rate.

\begin{figure}[htbp]
\vspace{0.5cm}
\par
\begin{center}
\includegraphics[width=0.5 \textwidth]{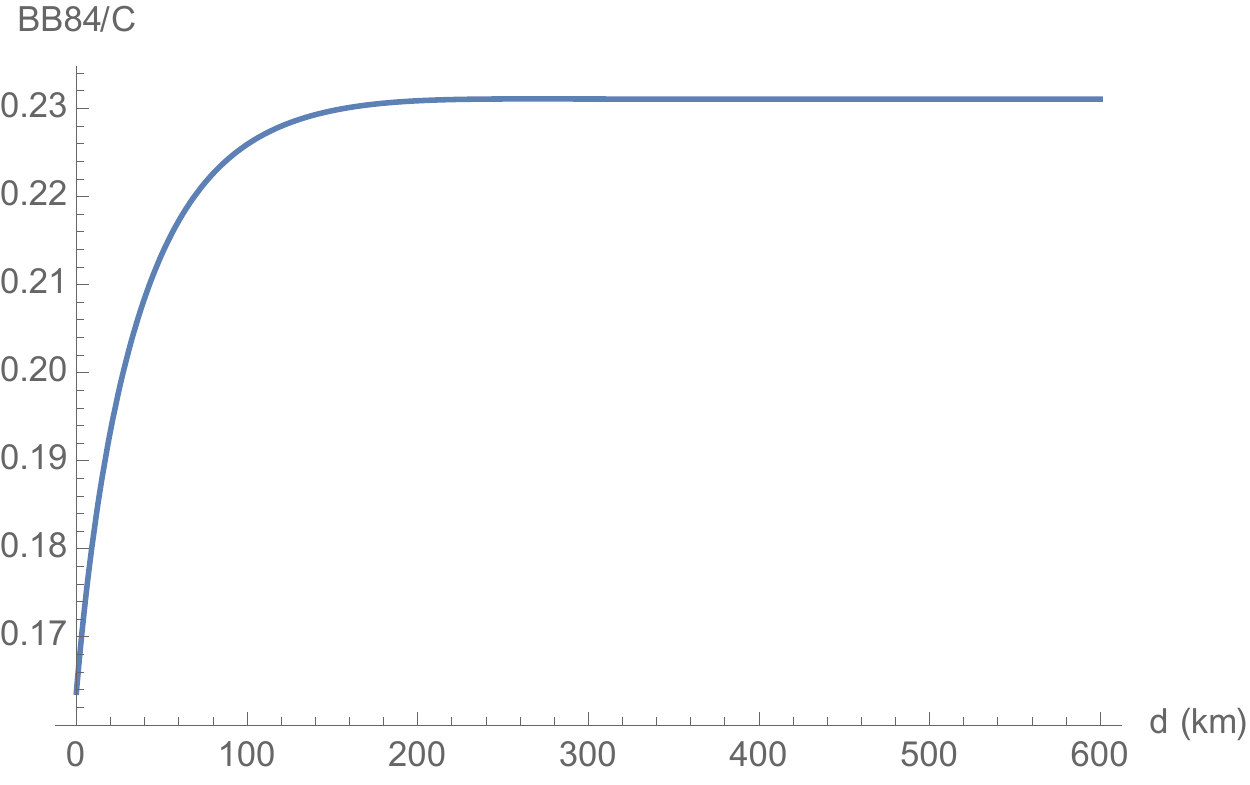} \vspace{-0.6cm}
\end{center}
\caption{Ratio of the secret key rate for an BB84 protocol using Rubidium memories taken from Ref.~\cite{vanLoock:2020gt} with the lossy-repeater capacity of Eq.~(\ref{chainCap1}). Parameters as per Fig.~\ref{Fig:rubidium}.}
\label{Fig:BB84comp}
\end{figure}

\section{Extension to general quantum networks}

Here, we extend the previous analysis from a linear to a more complex quantum network featuring an arbitrary topology, where the two end users aim at sharing entanglement or secret keys through single or multi-path routing strategies. 


\subsection{Preliminaries}

A \emph{quantum communication network} $\mathbf N$ involving $N$ nodes 
that can be interpreted as entities pursuing quantum communication can be described as an undirected graph $G=(V,E)$, where $V$ is the set of vertices or nodes ($|V|=N$), and $E$ the set of edges linking the elements in $V$. The set $E$ is determined by the underlying network infrastructure, i.e., an edge $(\nu_i,\nu_j)$ is an 
element of $E$ if there is a communication channel connecting the two  vertices $\nu_i$ and $\nu_j$. In 
a quantum communication scenario the nodes are linked together through a 
quantum channel $\mathcal E_{\nu_i-\nu_j}$. The transmission of quantum information through the quantum channel can be either forward $\nu_i\rightarrow\nu_j$ or backward $\nu_j\rightarrow\nu_i$. In what follows, we assign an orientation to the network so the quantum systems are always transmitted from sender $\nu_0$ to receiver $\nu_N$. 
This is a basic formalization of what is commonly called a quantum network.


Quantum information and entanglement can be transmitted and distributed along the network through a generic route $R$ between the two end-users, which is determined by the sequence of vertices $R=\nu_0-\cdots-\nu_i-\cdots-\nu_N$. In a single network $\mathbf N$, the different routes form a set $\mathbf R_{\mathbf N}=\{R_1,R_2,\ldots\}$. For each route there is an associated sequence of quantum channels, those involved in the routing process. As an example, in panel $a)$ of Fig.~\ref{Fig:network}, we show a fully-connected graph of four vertices that represents a diamond network. The set of all the possible routes from $\nu_0$ to $\nu_3$ is given by
$
\mathbf R_\diamond=\{R_1=\nu_0-\nu_1-\nu_3, R_2=\nu_0-\nu_2-\nu_3, R_3=\nu_0-\nu_1-\nu_2-\nu_3, R_4=\nu_0-\nu_2-\nu_1-\nu_3\}.
$

\subsection{Node-splitting in the network}

As we have done for the linear network, in order to account for a loss model for the stations, we proceed by splitting the nodes $\nu_i$ of the network and by inserting two quantum channels $\mathcal E_{\nu_i^1-\nu_i^2}$ and $\mathcal E_{\nu_i^2-\nu_i^3}$, connecting the three children nodes $\{\nu_i^1,\nu_i^2,\nu_i^3\}$. By doing so, the original network $\mathbf N$, described by the graph $G$, is mapped into $\mathbf N^\prime$ whose associated new graph is given by $G^\prime=(V^\prime,E^\prime)$, where $|V^\prime|=3N$.
\begin{figure}[h]
\vspace{0.2cm}
\par
\begin{center}
\includegraphics[width=0.49 \textwidth]{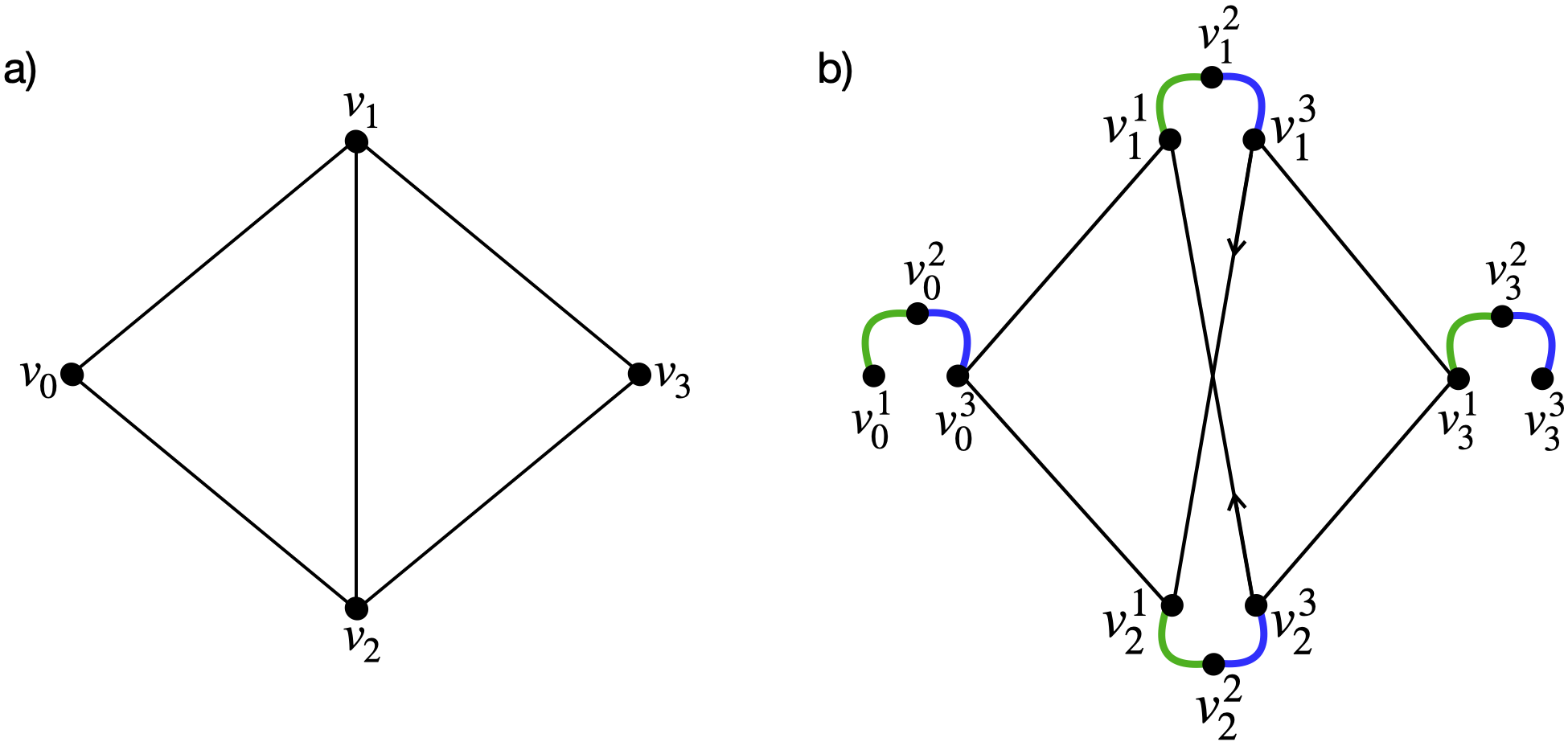} \vspace{-0.6cm}
\end{center}
\caption{A diamond network $\mathbf N$ of ideal nodes ($a$) is mapped into a network $\mathbf N^\prime$ of lossy nodes ($b$) by means of splitting.  Node $\nu_i$ is split in three children $\{\nu_i^1,\nu_i^2,\nu_i^3\}$ which are linked by additional edges $(\nu_i^1,\nu_i^2)$ and $(\nu_i^2,\nu_i^3)$ with associated lossy channels $\mathcal E_{\nu_i^1-\nu_i^2}$ and $\mathcal E_{\nu_i^2-\nu_i^3}$. The undirected link $(\nu_1,\nu_2)\in E$ in $\mathbf N$  is replaced, in $\mathbf N^\prime$, by two oriented links $\{(\nu_1^3,\nu_2^1),(\nu_2^3,\nu_1^1)\}\in E^\prime$.
 Accordingly, via the node-splitting, the route set $\mathbf R_\diamond$ is mapped into the route set $\mathbf R_\diamond^\prime$.}
\label{Fig:network}
\vspace{0.2cm}
\end{figure} 
The generic route $R$ of the ideal repeater network is updated to the route $R^\prime=\stackrel{\curvearrowright}{\nu_0}-\cdots-\stackrel{\curvearrowright}{\nu_i}-\cdots-\stackrel{\curvearrowright}{\nu_N}$, where we have defined the node internal route $\stackrel{\curvearrowright}{\nu_i}:=\nu_i^1-\nu_i^2-\nu_i^3$.
In panel $b)$ of Fig.~\ref{Fig:network} we show the node-splitting for the diamond network. 

It is important to note that, any edge belonging to two different routes with two opposite orientations, must be replaced by two distinct edges through node-splitting. More specifically, in the diamond network scenario of Fig.~\ref{Fig:network} by observing the route set $\mathbf R_{\diamond}$, the link connecting nodes $\nu_1$ and $\nu_2$ has two opposite orientation in route $R_3$ and route $R_4$. This means that, after node-splitting $\mathbf N\rightarrow\mathbf N^\prime$, the edge $(\nu_1,\nu_2)$ is replaced by two edges $(\nu_1^3,\nu_2^1)$ and $(\nu_2^3,\nu_1^1)$ with opposite orientations and the same associated quantum channel, i.e. $\mathcal E_{\nu_1^{3}-\nu_2^{1}}=\mathcal E_{\nu_2^{3}-\nu_1^{1}}$. These two links belong to the two distinct routes $R_3^\prime$ and $R_4^\prime$ of the new route set $\mathbf R_{\diamond}^\prime:=\{R^\prime_1=\stackrel{\curvearrowright}{\nu_0}-\stackrel{\curvearrowright}{\nu_1}-\stackrel{\curvearrowright}{\nu_3}, R^\prime_2=\stackrel{\curvearrowright}{\nu_0}-\stackrel{\curvearrowright}{\nu_2}-\stackrel{\curvearrowright}{\nu_3}, R^\prime_3=\stackrel{\curvearrowright}{\nu_0}-\stackrel{\curvearrowright}{\nu_1}-\stackrel{\curvearrowright}{\nu_2}-\stackrel{\curvearrowright}{\nu_3}, R^\prime_4=\stackrel{\curvearrowright}{\nu_0}-\stackrel{\curvearrowright}{\nu_2}-\stackrel{\curvearrowright}{\nu_1}-\stackrel{\curvearrowright}{\nu_3}\}$.
\begin{figure}[htbp]
\vspace{0.1cm}
\par
\begin{center}
\includegraphics[width=0.45 \textwidth]{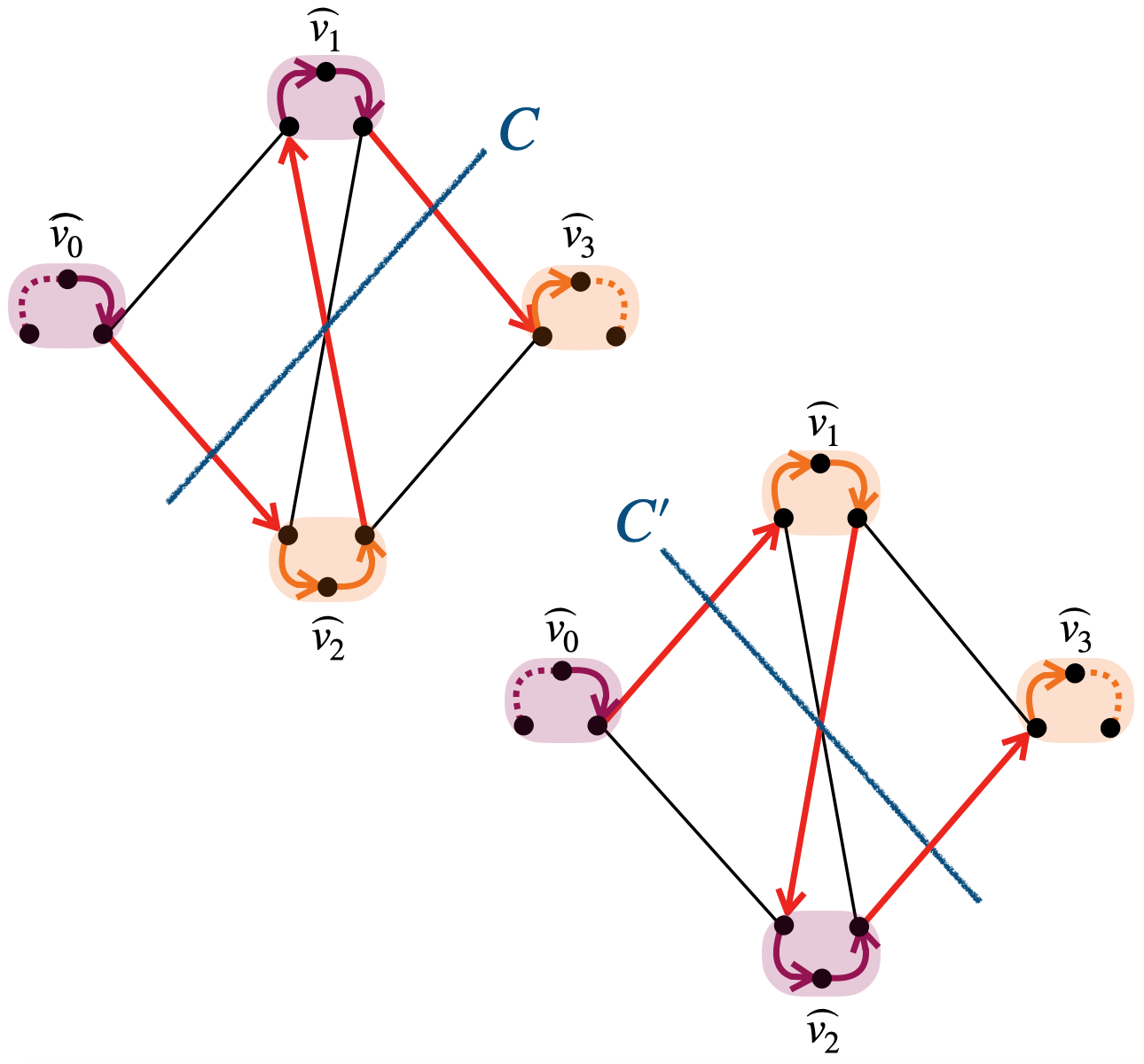} \vspace{-0.5cm}
\end{center}
\caption{Two examples of entanglement cut in a quantum diamond network of lossy nodes. The set of vertices $E^\prime$ of the network is divided into the two bipartitions $(V_A,V_B)$ and $(V_A^\prime,V_B^\prime)$ by the cuts $C$ and $C^\prime$ respectively. In the top network, $V_A=\{\stackrel{\frown}{\nu_0},\stackrel{\frown}{\nu_1}\}$ (purple), while $V_B=\{\stackrel{\frown}{\nu_2},\stackrel{\frown}{\nu_3}\}$ (orange). In the bottom network, $V_A^\prime=\{\stackrel{\frown}{\nu_0},\stackrel{\frown}{\nu_2}\}$ (purple), while $V_B^\prime=\{\stackrel{\frown}{\nu_1},\stackrel{\frown}{\nu_3}\}$ (orange). The induced cut sets (thick colored arrows) are respectively given by $K=\{(\stackrel{\frown}{\nu_0},\stackrel{\frown}{\nu_2}), (\stackrel{\frown}{\nu_2},\stackrel{\frown}{\nu_1}), (\stackrel{\frown}{\nu_1},\stackrel{\frown}{\nu_3}\}$ and $K^\prime=\{(\stackrel{\frown}{\nu_0},\stackrel{\frown}{\nu_1}), (\stackrel{\frown}{\nu_1},\stackrel{\frown}{\nu_2}), (\stackrel{\frown}{\nu_2},\stackrel{\frown}{\nu_3}\}$.}
\label{Fig:netcut}
\end{figure}


\subsection{Cuts of the lossy-repeater network}

An essential ingredient for our derivation is represented by the entanglement cut of the quantum network~\cite{PirNetwork}. 
Given the two end-nodes of a network of lossy repeaters, $\stackrel{\frown}{\nu_0}$ and $\stackrel{\frown}{\nu_N}$ (where $\stackrel{\frown}{\nu_i}:=\{\nu_i^1,\nu_i^2,\nu_i^3\}$), such an entanglement cut $C$ is defined as a bipartition $(V_A,V_B)$ of the set of nodes of the network such that  $\stackrel{\frown}{\nu_0}$ belongs to $V_A$ and $\stackrel{\frown}{\nu_N}$ belongs to $V_B$, with the elements of $V_A$ disconnected from the elements of $V_B$.
The entanglement cut induces the definition of the associated cut set $K$ which is the set of the links disconnected by the cut $C$. In Fig.~\ref{Fig:netcut}, we show two possible entanglement cuts of the diamond network in the presence of lossy repeater nodes. While the cut $C$ is always performed over the network link of the kind $(\nu_i^{3},\nu_j^1)$ between two distinct nodes $i$ and $j$, in the cut set $K$, we include also the internal repeater links which have vertices in common with the link disconnected by $C$. In other words, if $(\nu_i^{3},\nu_j^1)$ is a network link cut by $C$, the overall link $(\stackrel{\frown}{\nu_i},\stackrel{\frown}{\nu_j}):=(\nu_i^2,\nu_j^2)=(\nu_i^2,\nu_i^3)\cup(\nu_i^3,\nu_j^1)\cup(\nu_j^1,\nu_j^2)$ is an element of the cut set $K$, i.e. $K=\{(\stackrel{\frown}{\nu_i},\stackrel{\frown}{\nu_j})|,\stackrel{\frown}{\nu_i}\in V_A,\stackrel{\frown}{\nu_j}\in V_B\}$. Accordingly, the quantum channel associated to the generic element of the cut set is given by 
\begin{equation}
\mathfrak E_{i,j}:=\mathcal E_{\nu_j^1-\nu_j^{2}}\circ\mathcal E_{\nu_i^3-\nu_j^1}\circ\mathcal E_{\nu_i^{2}-\nu_i^{3}}~,
\label{netCh}
\end{equation}
and we set $\mathcal E_{\nu_0^1-\nu_0^2}=\mathcal E_{\nu_N^2-\nu_N^3}=\mathcal I$ for the two end nodes (see the dashed links in Fig.~\ref{Fig:netcut}).

\subsection{Single-path capacity of the lossy-repeater network}
Now that we have obtained a formalisation of the entanglement cuts for a lossy-repeater network (accounting for the node splitting), we are able to derive a corresponding formula for the single-path routing capacity. As per linear networks, our derivation is based on a straightforward generalisation of the ideal scenario with fully error-corrected repeaters. We know from Ref.~\cite[Th. 6 and 7]{PirNetwork} that the single-path capacity of a quantum network $\mathbf N$ of ideal-repeaters is bounded as follows
\begin{equation}\label{upper}
\mathcal C(\mathbf N)\leq\min_{C}E_R(C),
\end{equation}
where the right hand side term represents the minimization over all the possible cuts of the network of the single-path REE $E_R(C)$ associated to cut $C$. The latter quantity is defined by maximizing the REE over the edges of the cut set $K$, i.e.,
\begin{equation}
E_R(C):=\max_{(\nu_i,\nu_j)\in K}E_R(\rho_{\mathcal E_{i,j}}),
\end{equation}
where $\rho_{\mathcal E_{i,j}}$ is the Choi matrix of the lossy channel associated to the link $(\nu_i,\nu_j)$ (more technically this state and the associated REE are implicitly defined via asymptotic limits~\cite{PLOB}).

In contrast, a lower bound can be derived by finding the widest path in the quantum network~\cite{PirNetwork}, so we can write
\begin{equation}\label{lower}
\mathcal C(\mathbf N)\geq C(R^\star)=\min_C\mathcal C(C)
\end{equation}
where $R^\star$ is the optimal route such that the capacity of a single route $\mathcal C(R):=\min_\alpha \mathcal C(\mathcal E_\alpha^R)$ is maximum. Here $\alpha$ is the index over the route and we are implicitly defining $\mathcal E_\alpha^R:=\mathcal E_{\nu_i-\nu_j}$, with edge $(\nu_i,\nu_j)\in R$. Similarly, $\mathcal C(C):=\max_{(\nu_i,\nu_j)\in K}\mathcal C(\mathcal E_{\nu_i-\nu_j})$ is the single-path capacity associated to the cut. Furthermore, for a network of distillable channels, Eq.~(\ref{upper}) exactly coincides with Eq.~(\ref{lower}), and we can write~\cite{PirNetwork}
\begin{equation}\label{chainCap}
\mathcal C(\mathbf N)=\mathcal C(R^\star)=\min_C\mathcal C(C)=\min_{C}E_R(C).  \end{equation}

Thanks to the extension of the definition of the entanglement cut to the lossy repeater scenario, we can still rely on the chain of equalities in Eq.~(\ref{chainCap}).
Using the quantum channel defined in Eq.~(\ref{netCh}), we can therefore define the capacity of the single route $R^\prime\in\mathbf R^\prime_{\mathbf N^\prime}$ in the lossy-repeater network $\mathbf N^\prime$ as
\begin{equation}\label{routecap}
\mathcal C(R^\prime):=\min_{(\nu_i^{2},\nu_j^{2})\in R^\prime}\mathcal C(\mathfrak E_{i,j}^{R^\prime}).
\end{equation}
%
We notice that the links $(\nu_0^1,\nu_0^2)$ and $(\nu_N^2,\nu_N^3)$ belong to any possible existing single-path route of the lossy-repeater network, but since in our model they are both associated to a noiseless quantum channel, they can be disregarded in the definition of the route capacity. 

The main aim of our investigation is the analysis to the fundamental example of optical networks, where the link $(\nu_i^{3},\nu_j^1)$ connecting different nodes is described by a lossy channel with transmissivity $\eta_{i,j}$. We again assume that the two distinct quantum channels associated to the two internal repeater links $(\nu_i^1,\nu_i^2)$ and $(\nu_i^2,\nu_i^3)$ are represented by two lossy channels $\mathcal E_{\nu_i^1-\nu_i^2}$ and $\mathcal E_{\nu_i^2-\nu_i^3}$  with respective transmissivities $r_i$ and $t_i$. As a consequence, the quantum channel $\mathfrak E_{i,j}$, describing the effect of the transmission over the generic node-fibre-node link $(\nu_i^{2},\nu_j^{2})$, is a lossy channel with a transmissivity given by the product of the transmissivities of the involved lossy channels, i.e. $\mathcal T_{i,j}:=\eta_{i,j}r_it_j$ and capacity $\mathcal C(\mathfrak E_{i,j})=-\log_2(1-\mathcal T_{i,j})$. 

It then follows that the generic route $R^\prime\in\mathbf R^\prime_{\mathbf N^\prime}$ is identified by a collection of lossy channels with transmissivities $\{\mathcal T_{i,j}^{R^\prime}\}$. By defining the transmissivity of route $R^\prime$ as 
\begin{equation}\widetilde{\mathcal T}^{R^\prime}:=\min_{(\nu_i^{2},\nu_j^{2})\in R^\prime}\mathcal T_{i,j}^{R^\prime},
\end{equation}
its capacity reads
\begin{equation}\label{routeCap}
\mathcal C(R^\prime)= -\log_2(1-\widetilde{\mathcal T}^{R^\prime}).
\end{equation}
If we now maximize the expression in Eq.~(\ref{routeCap}) over the route set $\mathbf R^\prime_{\mathbf N^\prime}$, we obtain the single-path capacity of the lossy-repeater quantum network
\begin{align}\label{netCap1}
\mathcal C_{\text{loss}}(\mathbf N^\prime)&:=\max_{R^\prime\in\mathbf R^\prime_{\mathbf N^\prime}}\mathcal C(R^\prime)= -\log_2(1-\mathcal T),\\
\mathcal T&:=\max_{R^\prime\in\mathbf{R^\prime_{\mathbf N^\prime}}}\widetilde{\mathcal T}^{R^\prime}.
\end{align}
Equivalently, following the last terms of Eq.~(\ref{chainCap}), we can compute the capacity by minimizing, over all the possible cuts $C$, either the capacity of an entanglement cut $\mathcal C(C)$ or the REE of an entanglement cut $E_R(C)$. Thus, we may consider
\begin{align}
E_R(C)&:=\max_{(\nu_i^{2},\nu_j^{2})\in K}E_R(\rho_{\mathfrak E_{i,j}})\\
&=\max_{(\nu_i^{2},\nu_j^{2})\in K}[-\log_2(1-\mathcal T_{i,j})]
\nonumber
\\&=-\log_2(1-\widetilde{\mathcal T}_C),\nonumber
\end{align}
%
%
%
%
%
with $\widetilde{\mathcal T}_C=\max_{(\nu_i^{2},\nu_j^{2})\in K}\mathcal T_{i,j}$. We then obtain the single-path capacity of the lossy-repeater network via the minimization
\begin{equation}\label{netCap2}
\mathcal C_{\text{loss}}(\mathbf N^\prime)=\min_C[-\log_2(1-\widetilde{\mathcal T}_C)].
\end{equation}
By specifying Eqs.~(\ref{netCap1}) and~(\ref{netCap2}) to identical repeaters, i.e. $r_i=r_j=r$ and $t_i=t_j=t$, $\forall i,j=0,\ldots,|V|$, we get
\begin{equation}
\mathcal C_{\text{loss}}(\mathbf N^\prime)=-\log_2[(1-v\cdot\eta_{\mathbf N^\prime})],
\label{sLossCap}
\end{equation}
where we have defined $v:=rt$ and
\begin{equation}
\eta_{\mathbf{N^\prime}}:=\max_{R^\prime\in\mathbf{R^\prime_{\mathbf N^\prime}}}\min_{(\nu_i^{2},\nu_j^{2})}\eta_{i,j}^{R^\prime}=\min_C\max_{(\nu_i^{2},\nu_j^{2})\in K}\eta_{i,j}.
\end{equation}
The expressions above generalize the single-path capacity formulas of Ref.~\cite{PirNetwork} from ideal to lossy repeaters.





\subsection{Multi-path capacity of the lossy-repeater network}

A powerful routing strategy in a network is represented by flooding, where systems are transmitted in parallel so that each edge is exploited in each network use. 
Let us consider a quantum network $\mathbf N^\prime$ obtained, as described in the previous section, after node splitting $\mathbf N\rightarrow\mathbf N^\prime$, with a corresponding graph $G^\prime=(V^\prime,E^\prime)$ where $V^\prime=\{(\nu_i^1,\nu_i^2,\nu_i^3)\}_{i=0,\cdots,N}$. Once an orientation to the network $\mathbf N^\prime$ has been assigned, a multi-path flooding protocol can be defined as a collection of multicasts, each one realizing a point-to-multipoint communication. An orientation to the undirected network $\mathbf N$ is assigned by setting Alice ($\stackrel{\frown}{\nu_0}$) and Bob ($\stackrel{\frown}{\nu_N}$) respectively as the source and the sink of the network, and then by assigning a source-sink orientation to each edge of the network. Namely, for a generic link between the $i$-th and the $j$-th, we always identify $\nu_i^3$ as the source and $\nu_j^1$ as the sink. In this way a point-to-multipoint communication from node $\stackrel{\frown}{\nu_i}$ is defined as a quantum communication between $\stackrel{\frown}{\nu_i}$ and its out-neighborhood $D^{\text{out}}_{\stackrel{\frown}{\nu_i}}:=\{\nu_j^1\in V^\prime|(\nu_i^3,\nu_j^1)\in E_D^\prime\}$, with $E_D^\prime$ the edge-set $E^\prime$ where each element is now oriented. After the internal route $\stackrel{\curvearrowright}{\nu_0}$ at the sender's repeater station, the multi-path protocol starts with node $\nu_0^3$ sending quantum systems to each repeater station belonging to its neighbourhood.

The converse upper bound for the multi-path capacity $\tilde{\mathcal C}(\mathbf N)$ of a quantum network $\mathbf N$ is given by~\cite{PirNetwork}
\begin{equation}\label{multiUP}
\tilde{\mathcal C}(\mathbf N)\leq\min_C\tilde{E}_R(C),
\end{equation}
where the minimization is over all the possible cuts of the network and $\tilde{E}_R(C)$ is the multi-path REE flowing through an entanglement cut $C$. This is defined as the total REE of the cut set $K$ associated to $C$, namely
\begin{equation}
\tilde{E}_R(C):=\sum_{(\nu_i,\nu_j)\in K}E_R(\rho_{\mathcal E_{i,j}})~.
\end{equation}
An achievable rate (lower bound) for the multi-path capacity of the network is computed by applying the max-flow/min-cut theorem to the network, leading to~\cite{PirNetwork} 
\begin{equation}\label{multiLOW}
\tilde{\mathcal C}(\mathbf N)\geq\min_C\tilde{\mathcal C}(C)~,
\end{equation}
where $\mathcal C(C)$ is the multi-path capacity of an entanglement cut, defined by
\begin{equation}
\tilde{\mathcal C}(C):=\sum_{(\nu_i,\nu_j)\in K}\mathcal C(\mathcal E_{\nu_i-\nu_j})~.
\label{multicapcut}
\end{equation}

When the quantum network is connected by distillable quantum channels~\cite{PLOB}, the previous upper (\ref{multiUP}) and lower bound (\ref{multiLOW}) coincide and the multi-path capacity satisfies the following chain of equalities 
\begin{equation}
\tilde{\mathcal C}(\mathbf N)=\min_C\tilde{\mathcal C}(C)=\min_C\tilde{E}_R(C).
\end{equation}
Again we are able to generalize the analytical formulas, by extending the multi-path capacity for quantum and private communication over a quantum network from ideal to imperfect lossy nodes. For the fundamental case of an optical network connected by lossy channels (e.g., fibres), the crucial decomposition is the one in Eq.~(\ref{netCh}), where all the channels involved are lossy channels and therefore distillable.

Combining our decomposition with Eq.~(\ref{multicapcut}), we compute the multi-path capacity of an entanglement cut by summing up over the capacities of the quantum channels $\mathfrak E_{i,j}$ associated with the cut set $K$. We then have
\begin{align}
\tilde{\mathcal C}_{\text{loss}}(C)&=\sum_{(\stackrel{\frown}{\nu_i},\stackrel{\frown}{\nu_j})\in K}\mathcal C(\mathfrak E_{i,j})\\
&=\sum_{(\stackrel{\frown}{\nu_i},\stackrel{\frown}{\nu_j})\in K}E_R(\rho_{\mathfrak E_{i.j}})
\nonumber
\\
&=\sum_{(\stackrel{\frown}{\nu_i},\stackrel{\frown}{\nu_j})\in K}-\log_2(1-\mathcal T_{i,j})
\nonumber\\
&=-\log_2(L_C)
\nonumber
\end{align} 
where we have defined the total losses over a cut set as the product of the losses over the channels (repeater and link losses) in the cut set, i.e.
\begin{equation}
    L_C:=\prod_{(\stackrel{\frown}{\nu_i},\stackrel{\frown}{\nu_j})\in K}(1-\mathcal T_{i,j}).
\end{equation} 

Then the multi-path capacity of the quantum network with lossy repeaters 
is given by the minimization over all the possible entanglement cut of the above expression, i.e.,
\begin{align}
\tilde{\mathcal C}_{\text{loss}}(\mathbf N^\prime)&=\min_C\tilde{\mathcal C}_{\text{loss}}(C)\\
&=-\log_2(\max_CL_C)~.
\label{mLossCap}
\end{align}
It is easy to see that multi-path strategy is advantageous with respect to single-path even in the presence of lossy repeaters. For this purpose we can consider a split network $\mathbf N^\prime$ with identical repeaters (i.e. same loss) at each node and where all the network links $(\nu_i^3,\nu_j^1)$ are identical lossy channels with transmissivity $\eta$. Then from Eqs.~(\ref{sLossCap}) and (\ref{mLossCap}),
we get 
\begin{equation}
\tilde{\mathcal C}_{\text{loss}}(\mathbf N^\prime)=-\log_2(1-v\eta)^m=m\mathcal C_{\text{loss}}(\mathbf N^\prime),    
\end{equation}
where $m$ is the number of network links of the smallest allowed cut set. For instance, in the diamond network $\mathbf N_\diamond^\prime$ of Fig.~\ref{Fig:network} panel $b)$, the value of $m$ is equal to $2$.

\section{Conclusion and outlook}
Our work establishes analytical formulas for the maximum achievable rate of quantum and private communication between two end-users of a quantum network where the nodes are affected by internal loss.
In the linear 
\je{repeater chain}
scenario, we exploit a classical network technique, known as node splitting, to model the inevitable internal repeater loss.
In this way, we are able to describe the repeater chain as a suitable collection of distillable quantum channels, i.e. channels for which the lower and the upper bounds on the two-way assisted quantum (and private)  capacity coincide. 

Given this setting, by employing 
the powerful methodology of channel simulation and teleportation stretching, we have established an exact expression for the lossy-repeater capacity for quantum communication over a network with arbitrary number of lossy repeaters connected by pure-loss channels. Interestingly, when the number of repeaters increases, the derived capacity turns out to be a function of the internal loss of a single node, which then acts as the ultimate upper limit to the maximum achievable rate for quantum and private communication. 

Finally, we have considered the important role played by \emph{time} that must be taken into account in any actual implementation of a quantum repeater chain. In such a realistic setting, we have shown how the performance can indeed
overcome the repeaterless PLOB bound and approach the optimal single-repeater bound, even in the presence of internal time-dependent loss, e.g., induced by limited coherence times. 

The present study can be seen as a relevant step in an 
important direction and invites further studies in many ways. This work has put an emphasis on losses, which in most practical implementations is indeed the main source of errors. 
A more detailed study should accommodate dark counts and further offset noise as well. On a broader level, the work hopes to push forward
a line of thought aiming at identifying the ultimate bounds for practically achievable rates in quantum repeater schemes, capturing high-level distinctions between protocols without going too much into specifics of a particular implementation. Such considerations, it is reasonable to expect, will substantially help in assessing the potential of multi-partite long-distance quantum communication.

\section{Acknowledgments}
J.~E.~has been supported by the BMBF (QR.X on quantum repeaters) and the Einstein Foundation. R.~L.~acknowledges support from the Alexander von Humboldt Foundation. N.~W.~has been funded by a Marie Sklodowska-Curie Individual Fellowship. S.~P.~has been supported by the European Union via ``Continuous Variable Quantum Communications'' (CiViQ, Grant Agreement No.~820466). The authors would like to thank Frederik Hahn, Julius Walln{\"o}fer and Peter van Loock for interesting discussions. We are also grateful to Cillian Harney for his feedback that helped us improve Fig.~\ref{Fig:rubidium}.

\section*{Appendix}
\subsection*{Appendix A: Two-way quantum capacities and general bounds}
The most important point-to-point quantum communication scenario concerns two remote parties, Alice and Bob, which are connected by a (memoryless) quantum channel $\mathcal E$ without pre-sharing any entanglement. By means of this channel, the two parties may implement various quantum tasks as for instance the reliable transmission of qubits, the distillation of entanglement bits (ebits) and the generation of secret bits. The most general protocols are based on transmissions through the quantum channel which are interleaved by local operations (LO) assisted by unlimited and two-way classical communication (CC), briefly called adaptive LOCCs.
At the beginning of this protocol, Alice and Bob have two local registers $a$ and $b$ of quantum systems which are adaptively updated before and after each transmission through $\mathcal E$. After a number $n$ of channel's uses, Alice and Bob will share the quantum state $\rho_{a,b}^n$ which depends on the sequence of LOCCs $\mathcal L=\{L_1,L_2,\cdots,L_n\}$.

The rate $R_n$ of this protocol is defined through a target state $\phi_n$ whose content of information is equal to $nR_n$ bits. If the output state $\rho_{a,b}^n$ is close in the trace norm to $\phi_n$, i.e. $\|\rho_{a,b}^n-\phi_n\|\leq\epsilon$ for some $\epsilon\rightarrow>0$, 
then the rate of the protocol is equal to $R_n$. The generic two-way capacity $\mathcal C(\mathcal E)$ is defined by taking the limit for a large number of channel's uses $n$ and by optimizing over all the possible adaptive protocols $\mathcal L$, i.e.
\begin{equation}
\mathcal C(\mathcal E):=\sup_{\mathcal L}\lim_{n\rightarrow\infty}R_n~.\label{ChCAP}   
\end{equation}

In order for the quantity $\mathcal C(\mathcal E)$ to get an operational meaning, we need to specify the goal of the adaptive protocol implemented by Alice and Bob. Thus, if the target state is a maximally entangled state, meaning that the protocol is an entanglement distribution protocol, we have that $\mathcal C(\mathcal E)=D_2(\mathcal E)$, where $D_2(\mathcal E)$ is the two-way entanglement distribution capacity of the channel. Since an ebit can teleport a qubit and viceversa with a qubit is possible distribute an ebit, $D_2(\mathcal E)$ is equal to the two-way quantum capacity $Q_2(\mathcal E)$, i.e. the maximum achievable rate for transmitting quantum information. If the protocol is a QKD protocol, $\phi_n$ is a private state and the generic two-way quantum capacity is the secret key capacity $K(\mathcal E)$ which is equal to the private capacity $P_2(\mathcal E)$ (unlimited two-way CCs and one time-pad). Since a maximally entangled state is a specific type of
private state, we can write the following relations between all the different capacities
\begin{equation}\label{capIneq}
Q_2(\mathcal E)=D_2(\mathcal E)\leq P_2(\mathcal E)=K(\mathcal E)~.
\end{equation}

As one can see from Eq.~(\ref{ChCAP}), the quantity $\mathcal C(\mathcal E)$ cannot be evaluated directly from its definition and the best strategy to assess it is to resort to suitable lower and upper bounds that are usually built upon information and entanglement measures.

A general lower bound can be given in terms of the \emph{coherent}~\cite{Coh1,Coh2} or \emph{reverse coherent information}~\cite{RevCoh1,RevCoh2} which are, respectively, defined as
\begin{align}
I_C(\mathcal E,\rho_A)&=I(A\langle B)_{\rho_{RB}}:=S(\rho_B)-S(\rho_{RB}),\\ 
I_{RC}(\mathcal E,\rho_A)&=I(A\rangle B)_{\rho_{RB}}:=S(\rho_R)-S(\rho_{RB}),
\end{align}
where the quantum channel $\mathcal E$ takes as an input the quantum state $\rho_A$ of system $A$ (see also the related notions of 
\emph{negative cb-entropy} of a channel~\cite{devetakcb} and 
\emph{pseudo-coherent information}~\cite{hayashipseudo}). If $R$ is an auxiliary system and $|\psi\rangle_{RA}$ the purification of $\rho_A$, then the output of the channel is $\rho_{RB}=(\mathcal I\otimes\mathcal E)(|\psi\rangle\langle\psi|_{RA})$. 
%
%
In the above expressions, we also have $\rho_{R(B)}=\Tr_{B(R)}\rho_{RB}$ and $S(\rho):=-\Tr(\rho\log_2\rho)$ is the 
\emph{von Neumann entropy}. 
When the input state $\rho_A$ is a maximally-mixed state, its purification is a maximally-entangled state $\Phi_{RA}$, so that $\rho_{RB}$ becomes the Choi matrix of the channel $\sigma_{\mathcal E}=(\mathcal I\otimes\mathcal E) (\Phi_{RA})$. Then we can define the coherent and reverse coherent information of the quantum channel $\mathcal E$ respectively as follows~\cite[Supp. Note 2]{PLOB}
\begin{align}
I_C(\mathcal E)&:=I(A\langle B)_{\sigma_{\mathcal E}}\label{coherent},\\
I_{RC}(\mathcal E)&:=I(A\rangle B)_{\sigma_{\mathcal E}}.\label{Rcoherent}
\end{align}
The quantity $I_C(\mathcal E)$ constitutes an achievable rate for {\itshape forward} one-way entanglement distillation, whereas $I_{RC}(\mathcal E)$ is an achievable rate for {\itshape backward} one-way entanglement distillation. In fact, due to the 
\emph{hashing inequality}~\cite{DevWint}, 
we can write
\begin{equation}\label{lowerB}
\max\{I_C(\mathcal E),I_{RC}(\mathcal E)\}\leq D_1(\sigma_{\mathcal E})~,   
\end{equation}
where $D_1(\sigma_\mathcal E)$ is the entanglement that can be distilled from the channel's Choi matrix with the assistance of forward or backward classical communication.

The general weak converse upper bound to the two-way quantum capacity $\mathcal C(\mathcal E)$ derived in Ref.~\cite{PLOB}, is built upon the notion of the \emph{relative entropy of entanglement} (REE) 
\cite{RE2}
suitably extended from quantum 
states to quantum channels.
Let us recall that the REE of a 
quantum state $\rho$ is defined as the 
minimum relative entropy between $\rho$ and a 
separable state $\sigma_s$~\cite{RE1,RE2}, i.e.,
\begin{equation}
    E_R(\rho):=\inf_{\sigma_s\in\text{SEP}}S(\rho\|\sigma_s)~.
\end{equation}
We can also introduce the REE of a discrete variable quantum channel $\mathcal E$ with associated Choi matrix $\sigma_{\mathcal E}$ in the following way
\begin{equation}
E_R(\mathcal E):=\sup_{\rho}E_R[
(\mathcal I\otimes\mathcal E) (\rho)]\leq E_R(\sigma_{\mathcal E})~.    
\end{equation}
Then Ref.~\cite[Th.~1]{PLOB} states that generic two-way capacity of equation~(\ref{ChCAP}) is upper bounded by the REE bound 
\begin{equation}\label{upperB}
 \mathcal C(\mathcal E)\leq E_R^\star(\mathcal E):=\sup_{\mathcal L}\lim_{n\rightarrow\infty}\frac{E_R(\rho_{a,b}^n)}{n}~,  
\end{equation}
where $\rho_{a,b}^n$ is the output state of a $n$-use adaptive protocol $\mathcal L$.
Note that both the lower bound (\ref{lowerB}) and the upper bound (\ref{upperB}) hold for an arbitrary quantum channel in arbitrary dimension.
In the subsequent section, we discuss how to extend them to asymptotic states, providing in this way a formulation for CV systems, following the asymptotic methodology of Ref.~\cite{PLOB}.

\subsection*{Appendix B: Asymptotic formulation for bosonic systems}

It is important to note that when dealing with continuous variable systems the maximally entangled state is an asymptotic state (energy-unbounded) obtained as the limit $\Phi:=\lim_\mu\Phi^\mu$, where $\Phi^\mu$ is a sequence of two mode squeezed vacuum (TMSV) states parametrized by $\mu$ which quantifies both the two-mode squeezing and the mean number $\bar n$ of thermal photons (local energy) in both modes, i.e., $\mu=\bar n+1/2$~\cite{TeleBK,PeterRMP}. According to this, the Choi state of a bosonic channel $\mathcal E$ (e.g. the pure-loss channel under consideration) is given by the asymptotic limit
\begin{equation}
\sigma_{\mathcal E}:=\lim_\mu\sigma_{\mathcal E}^\mu,\quad\sigma_{\mathcal E}^\mu:=
(\mathcal I\otimes\mathcal E)(\Phi^\mu)~.\label{AsymChoi}
\end{equation}
Correspondingly the computation of the (reverse) coherent information of a quantum channel introduced in Eq.~(\ref{coherent}) and (\ref{Rcoherent}) has to be performed as the following limits
\begin{align}
&I(A\langle B)_{\sigma_{\mathcal E}}:=\lim_{\mu\rightarrow\infty}I(A\langle B)_{\sigma_{\mathcal E}^\mu},\\
&I(A\rangle B)_{\sigma_{\mathcal E}}:=\lim_{\mu\rightarrow\infty}I(A\rangle B)_{\sigma_{\mathcal E}^\mu}~.
\end{align}
For \emph{bosonic Gaussian channels}
\cite{GaussianChannel}, it can be shown that the functionals $I(A\langle B)_{\sigma_{\mathcal E}^\mu}$ and $I(A\rangle B)_{\sigma_{\mathcal E}^\mu}$ are continuous, monotonic and bounded
in $\mu$. Therefore, the above limits are finite and we can continuously extend Eq.~(\ref{lowerB}) to the asymptotic Choi matrix of a CV channel, for which we may set $D_1(\mathcal E):=\lim_{\mu\rightarrow\infty}D_1(\sigma_{\mathcal E}^\mu)$.\\ 
Let us now consider two sequences of states $\rho_1^\mu$ and $\rho_2^\mu$ converging, respectively, in the trace norm to $\rho_1$ and $\rho_2$, i.e., $\|\rho_i^\mu-\rho_i\|\rightarrow0$, for $i=1,2$. By exploiting the lower semi-continuity of the relative entropy, 
we can write 
\begin{equation}
S(\rho_1\|\rho_2)\leq\liminf_{\mu\rightarrow\infty}S(\rho_1^\mu\|\rho_2^\mu)~.
\end{equation}
As a consequence the relative entropy of entanglement of an asymptotic state $\rho=\lim_\mu\rho^\mu$ is defined as follows
\begin{equation}\label{asREE}
E_R(\rho):=\inf_{\rho_s^\mu}\liminf_{\mu\rightarrow\infty}S(\rho^\mu\|\rho^\mu_s)~,
\end{equation}
where $\rho_s^\mu$ is an arbitrary sequence of separable states satisfying $\|\rho_s^\mu-\rho_s\|\stackrel{\mu\rightarrow\infty}{\longrightarrow}0$ for some separable state $\rho_s$. A direct implication of Eq.~(\ref{asREE}) is that the REE computed over the {\itshape quasi}-Choi matrix $\sigma_{\mathcal E}^\mu$ of a bosonic channel is defined as
\begin{equation}
E_R(\sigma_{\mathcal E}):=\inf_{\rho_s^\mu}\liminf_{\mu\rightarrow+\infty}S(\sigma_{\mathcal E}^\mu\|\rho_s^\mu)
\end{equation}

\subsection*{Appendix C: Channel simulation and teleportation stretching}

We already mentioned that in order to write Eq.~(\ref{stretch}), which is fundamental in simplifying the REE bound of Eq.~(\ref{upperB}), we need to rely on two ingredients which are, respectively, known as {\itshape channel simulation} and {\itshape teleportation stretching}.
In this section we briefly review these two technical steps with the main definitions while referring the reader to \cite{PLOB} for more technical details and a discussion of historical developments.

The notion of quantum channel simulation comes from a straightforward generalization of quantum teleportation protocol whose structure involves local operations (LO), Bell detection on Alice’s side and Bob’s unitary correction, plus classical communication (CC) from Alice to Bob~\cite{BennettTELE}. For a maximally entangled resource state $\Phi$, the teleported output perfectly correspond to the input. If we perform teleportation over an arbitrary mixed resource state  of systems A and B, the teleported state on Bob’s side will result in the output of a certain quantum channel $\mathcal E$ from Alice to Bob (see Ref.~\cite[Sec.~V]{BDSW96} for the initial insights of this technique, later expanded by various groups over the years). 

More generally, any implementation through an arbitrary LOCC $\mathbb L$ and a resource state $\sigma$ simulates the output of a quantum channel $\mathcal E$.
Thus, for any $\mathcal E$ and for any input $\rho$, we can express the output as~\cite{PLOB}
\begin{equation}
    \mathcal E(\rho)=\mathbb L(\rho\otimes\sigma)~.\label{simula}
\end{equation}
 When dealing with CV systems as in our scenario, the LOCC simulation involves the limit $ \sigma:=\lim_{\mu\rightarrow\infty}\sigma^\mu$ of resource states $\sigma^\mu$. Then, for any finite $\mu$, the simulation provides the approximated channel
\begin{equation}
\mathcal E^\mu(\rho)=\mathbb L(\rho\otimes\sigma^\mu)~,
\end{equation}
which defines the quantum channel $\mathcal E$ as the following point-wise limit
\begin{equation}
\mathcal E(\rho)=\lim_{\mu\rightarrow\infty}\mathcal E^\mu(\rho)~.\label{LimSimu}
\end{equation}
For any given quantum channel, we can always find a suitable LOCC $\mathbb L$ and a resource state $\sigma$ that achieve the simulation in Eq.~(\ref{simula}). A genuine LOCC simulation is established when the quantum channel satisfies the property of {\itshape teleportation covariance}. If $\mathcal U$ is the group of teleportation unitaries, a quantum channel $\mathcal E$ is teleportation covariant if the following identity holds for any $U\in\mathcal U$
\begin{equation}
\mathcal E(U\rho U^\dagger)=V\mathcal E(\rho)V^\dagger~,
\end{equation}
with $V$ a unitary transformation not necessarily belonging to $\mathcal U$. Note that the unitary group $\mathcal U$ is the Weyl-Heisenberg group (generalized Pauli operators) for DV systems, while for CV systems it coincides with the group of displacement operators. An interesting property of a tele-covariant quantum channel $\mathcal E$ is that it can be simulated by teleporting the input state $\rho$ using its Choi matrix $\sigma_{\mathcal E}$ as the resource for teleportation, i.e., for a DV channel we write
\begin{equation}
\mathcal E(\rho)=\mathbb T(\rho\otimes\sigma_{\mathcal E}) 
\end{equation}
where $\mathbb T$ is teleportation~\cite{BennettTELE}. For a CV channel, by recalling Eq.~(\ref{LimSimu}), the above relation is rewritten as
\begin{equation}\label{Sim1}
\mathcal E^\mu(\rho)=\mathbb T(\rho\otimes\sigma^\mu_{\mathcal E})~,
\end{equation}
where now $\mathbb T$ is the Braunstein-Kimble teleportation~\cite{TeleBK,TeleportationReview} and the asymptotic Choi state $\sigma_{\mathcal E}^\mu$ defines the asymptotic Choi state for large $\mu$ as in Eq.~(\ref{AsymChoi}). Note that several quantum channels satisfy the property of teleportation covariance, including Pauli and erasure channels in DVs, and bosonic Gaussian channels in CVs.

By making use of channel simulation, we are able to perform teleportation stretching and to simplify the adaptive structure of a protocol for quantum and private communication. This means that the protocol output $\rho_{a,b}^n$ is reduced into an $n$-fold tensor product of resource states $\sigma^{\otimes n}$ up to a TP LOCC $\bar{\Lambda}$. The reduction procedure starts by replacing each transmission over the channel $\mathcal E$ with its simulation $(\mathbb T,\sigma)$. At this stage, we can then stretch the resource state $\sigma$ outside the adaptive operations, while $\mathbb T$ is incorporated into the protocol LOCCs. After that, all the LOCCs together with the initial register preparation, are merged into a single final LOCC $\bar\Lambda$, which turns out to be TP after averaging over all the possible local measurement outcomes. At the end we can then write~\cite[Lemma~3]{PLOB}
\begin{equation}\label{stretch1}
\rho_{a,b}^n=\bar\Lambda(\sigma^{\otimes n})~.    
\end{equation}

For CV quantum channels, the above equation must be interpreted in an asymptotic fashion in the following manner. We replace each transmission through $\mathcal E$ with the channel $\mathcal E^\mu$ defined in (\ref{Sim1}) with a finite-energy resource state $\sigma^\mu$. If we assume that the local registers of Alice and Bob have energy $\leq N$, i.e., the total input state of each transmission belongs to a bounded alphabet $D_N$, the channel $\mathcal E^\mu$ simulates $\mathcal E$ up to an error given by $\epsilon(\mu,N):=\|\mathcal E-\mathcal E^\mu\|_{\diamond N}$, where 
\begin{equation}
\|\mathcal E-\mathcal E^\prime\|_{\diamond N}:=\sum_{\rho_{RS}\in D_N}\|\mathcal I_R\otimes \mathcal E_S(\rho_{RS})-\mathcal I_R\otimes \mathcal E^\prime_S(\rho_{RS})\| 
\end{equation}
is the energy constrained diamond norm. By exploiting the non-increasing of the trace distance under CPTP maps and the triangle inequality, it can be proven~\cite{PLOB} that the trace distance between the output $\rho_{a,b}^n$ and the simulated output $\rho_{a,b}^{n,\mu}$ (the output of an adaptive protocol performed over $\mathcal E^\mu$) satisfies
\begin{equation}
\|\rho_{a,b}^n-\rho_{a,b}^{n,\mu}\|\leq n\epsilon(\mu,N)~.  \end{equation}
We can now substitute $\rho_{a,b}^{n,\mu}$ with its decomposition given by the teleportation stretching, so that we obtain
\begin{equation}
    \|\rho_{a,b}^n-\bar\Lambda(\sigma^{\mu\otimes n})\|\leq n\epsilon(\mu,N)~,
\end{equation}
for any energy constrain $N$. Then by taking the limit for $\mu\rightarrow\infty$ we get the asymptotic version of Eq.~(\ref{stretch}) (asymptotic stretching)
\begin{equation}\label{stretch2}
 \lim_{\mu\rightarrow\infty}\|\rho_{a,b}^n-\bar\Lambda(\sigma^{\mu\otimes n})\|=0~.
\end{equation}
By using the decompositions of Eq.~(\ref{stretch1}) and (\ref{stretch2}) we can consequently simplify the upper bound in (\ref{upper}). In fact we can write 
\begin{equation}\label{REEineq}
E_R(\rho_{a,b}^n)\leq E_R(\sigma^{\otimes n})\leq nE_R(\sigma)~,    
\end{equation}
where in the two inequalities the monotonicity of the REE under TP LOCCs and the sub-additivity of the REE over tensor products are, respectively, exploited. By putting Eq.~(\ref{REEineq}) into Eq.~(\ref{upper}), we can get rid of both the optimization over all the adaptive protocols and the asymptotic limit so that a {\itshape single-letter} upper bound to the capacities introduced in (\ref{capIneq}) is obtained
\begin{equation}
  \mathcal C(\mathcal E)\leq E_R(\sigma)~.  
\end{equation}
If the channel is teleportation covariant we can then write the above equation in terms of the Choi matrix $\sigma_{\mathcal E}$ of the channel, i.e.,
\begin{equation}
  \mathcal C(\mathcal E)\leq E_R(\sigma_{\mathcal E})~.
  \end{equation}
See also Ref.~\cite[Th.~5]{PLOB} and related proofs for more details.

\subsection*{Appendix D: BB84 key rate}

Over a pure loss channel there is no dephasing so there is one bit of distillable key for every successful connection between the two remote stations. All that is required then is to calculate the probability of this happening for a single channel use for a repeater scheme based upon storage and entanglement swapping, but without any distillation. Following Ref.~\cite{vanLoock:2020gt} we can calculate that, for a scheme with a half-link success probability, $p$, given by (\ref{psimp}) and symmetric transmission and receiver losses $\tau^{t,\mathrm{eff}} = \tau^{r,\mathrm{eff}} = \tau^{\mathrm{eff}}$ that the BB84 rate is
\eqn{r_{\mathrm{BB84}} = \frac{1}{2} \frac{ \sqrt{\eta}\tau^{\mathrm{eff}}(2-\sqrt{\eta}\tau^{\mathrm{eff}})}{3-2\sqrt{\eta}\tau^{\mathrm{eff}}} \tau_{\mathrm{mem}} .}
The rate per time is 
then given by $R r_{\mathrm{BB84}}$. For a standard polarisation based implementation, there are actually two
optical modes available (corresponding to horizontal and vertical polarisation) that must be transmitted for each round, so this rate must be halved to get the rate per transmitted mode, which gives the factor of $1/2$ in the above expression.

\begin{thebibliography}{47}%
\makeatletter
\providecommand \@ifxundefined [1]{%
 \@ifx{#1\undefined}
}%
\providecommand \@ifnum [1]{%
 \ifnum #1\expandafter \@firstoftwo
 \else \expandafter \@secondoftwo
 \fi
}%
\providecommand \@ifx [1]{%
 \ifx #1\expandafter \@firstoftwo
 \else \expandafter \@secondoftwo
 \fi
}%
\providecommand \natexlab [1]{#1}%
\providecommand \enquote  [1]{``#1''}%
\providecommand \bibnamefont  [1]{#1}%
\providecommand \bibfnamefont [1]{#1}%
\providecommand \citenamefont [1]{#1}%
\providecommand \href@noop [0]{\@secondoftwo}%
\providecommand \href [0]{\begingroup \@sanitize@url \@href}%
\providecommand \@href[1]{\@@startlink{#1}\@@href}%
\providecommand \@@href[1]{\endgroup#1\@@endlink}%
\providecommand \@sanitize@url [0]{\catcode `\\12\catcode `\$12\catcode
  `\&12\catcode `\#12\catcode `\^12\catcode `\_12\catcode `\%12\relax}%
\providecommand \@@startlink[1]{}%
\providecommand \@@endlink[0]{}%
\providecommand \url  [0]{\begingroup\@sanitize@url \@url }%
\providecommand \@url [1]{\endgroup\@href {#1}{\urlprefix }}%
\providecommand \urlprefix  [0]{URL }%
\providecommand \Eprint [0]{\href }%
\providecommand \doibase [0]{http://dx.doi.org/}%
\providecommand \selectlanguage [0]{\@gobble}%
\providecommand \bibinfo  [0]{\@secondoftwo}%
\providecommand \bibfield  [0]{\@secondoftwo}%
\providecommand \translation [1]{[#1]}%
\providecommand \BibitemOpen [0]{}%
\providecommand \bibitemStop [0]{}%
\providecommand \bibitemNoStop [0]{.\EOS\space}%
\providecommand \EOS [0]{\spacefactor3000\relax}%
\providecommand \BibitemShut  [1]{\csname bibitem#1\endcsname}%
\let\auto@bib@innerbib\@empty
\bibitem [{\citenamefont {Pirandola}\ \emph {et~al.}(2020)\citenamefont
  {Pirandola}, \citenamefont {Andersen}, \citenamefont {Banchi}, \citenamefont
  {Berta}, \citenamefont {Bunandar}, \citenamefont {Colbeck}, \citenamefont
  {Englund}, \citenamefont {Gehring}, \citenamefont {Lupo}, \citenamefont
  {Ottaviani}, \citenamefont {Pereira}, \citenamefont {Razavi}, \citenamefont
  {Shaari}, \citenamefont {Tomamichel}, \citenamefont {Usenko}, \citenamefont
  {Vallone}, \citenamefont {Villoresi},\ and\ \citenamefont
  {Wallden}}]{Pirandola:20}%
  \BibitemOpen
  \bibfield  {author} {\bibinfo {author} {\bibfnamefont {S.}~\bibnamefont
  {Pirandola}}, \bibinfo {author} {\bibfnamefont {U.~L.}\ \bibnamefont
  {Andersen}}, \bibinfo {author} {\bibfnamefont {L.}~\bibnamefont {Banchi}},
  \bibinfo {author} {\bibfnamefont {M.}~\bibnamefont {Berta}}, \bibinfo
  {author} {\bibfnamefont {D.}~\bibnamefont {Bunandar}}, \bibinfo {author}
  {\bibfnamefont {R.}~\bibnamefont {Colbeck}}, \bibinfo {author} {\bibfnamefont
  {D.}~\bibnamefont {Englund}}, \bibinfo {author} {\bibfnamefont
  {T.}~\bibnamefont {Gehring}}, \bibinfo {author} {\bibfnamefont
  {C.}~\bibnamefont {Lupo}}, \bibinfo {author} {\bibfnamefont {C.}~\bibnamefont
  {Ottaviani}}, \bibinfo {author} {\bibfnamefont {J.~L.}\ \bibnamefont
  {Pereira}}, \bibinfo {author} {\bibfnamefont {M.}~\bibnamefont {Razavi}},
  \bibinfo {author} {\bibfnamefont {J.~Shamsul}\ \bibnamefont {Shaari}},
  \bibinfo {author} {\bibfnamefont {M.}~\bibnamefont {Tomamichel}}, \bibinfo
  {author} {\bibfnamefont {V.~C.}\ \bibnamefont {Usenko}}, \bibinfo {author}
  {\bibfnamefont {G.}~\bibnamefont {Vallone}}, \bibinfo {author} {\bibfnamefont
  {P.}~\bibnamefont {Villoresi}}, \ and\ \bibinfo {author} {\bibfnamefont
  {P.}~\bibnamefont {Wallden}},\ }\bibfield  {title} {\enquote {\bibinfo
  {title} {Advances in quantum cryptography},}\ }\href {\doibase
  10.1364/AOP.361502} {\bibfield  {journal} {\bibinfo  {journal} {Adv. Opt.
  Photon.}\ }\textbf {\bibinfo {volume} {12}},\ \bibinfo {pages} {1012--1236}
  (\bibinfo {year} {2020})}\BibitemShut {NoStop}%
\bibitem [{\citenamefont {Xu}\ \emph {et~al.}(2020)\citenamefont {Xu},
  \citenamefont {Ma}, \citenamefont {Zhang}, \citenamefont {Lo},\ and\
  \citenamefont {Pan}}]{RevModPhys.92.025002}%
  \BibitemOpen
  \bibfield  {author} {\bibinfo {author} {\bibfnamefont {F.}~\bibnamefont
  {Xu}}, \bibinfo {author} {\bibfnamefont {X.}~\bibnamefont {Ma}}, \bibinfo
  {author} {\bibfnamefont {Q.}~\bibnamefont {Zhang}}, \bibinfo {author}
  {\bibfnamefont {H.-K.}\ \bibnamefont {Lo}}, \ and\ \bibinfo {author}
  {\bibfnamefont {J.-W.}\ \bibnamefont {Pan}},\ }\bibfield  {title} {\enquote
  {\bibinfo {title} {Secure quantum key distribution with realistic devices},}\
  }\href {\doibase 10.1103/RevModPhys.92.025002} {\bibfield  {journal}
  {\bibinfo  {journal} {Rev. Mod. Phys.}\ }\textbf {\bibinfo {volume} {92}},\
  \bibinfo {pages} {025002} (\bibinfo {year} {2020})}\BibitemShut {NoStop}%
\bibitem [{\citenamefont {Gisin}\ \emph {et~al.}(2002)\citenamefont {Gisin},
  \citenamefont {Ribordy}, \citenamefont {Tittel},\ and\ \citenamefont
  {Zbinden}}]{RevModPhys.74.145}%
  \BibitemOpen
  \bibfield  {author} {\bibinfo {author} {\bibfnamefont {N.}~\bibnamefont
  {Gisin}}, \bibinfo {author} {\bibfnamefont {G.}~\bibnamefont {Ribordy}},
  \bibinfo {author} {\bibfnamefont {W.}~\bibnamefont {Tittel}}, \ and\ \bibinfo
  {author} {\bibfnamefont {H.}~\bibnamefont {Zbinden}},\ }\bibfield  {title}
  {\enquote {\bibinfo {title} {Quantum cryptography},}\ }\href {\doibase
  10.1103/RevModPhys.74.145} {\bibfield  {journal} {\bibinfo  {journal} {Rev.
  Mod. Phys.}\ }\textbf {\bibinfo {volume} {74}},\ \bibinfo {pages} {145--195}
  (\bibinfo {year} {2002})}\BibitemShut {NoStop}%
\bibitem [{\citenamefont {Pirandola}\ \emph {et~al.}(2015)\citenamefont
  {Pirandola}, \citenamefont {Eisert}, \citenamefont {Weedbrook}, \citenamefont
  {Furusawa},\ and\ \citenamefont {Braunstein}}]{TeleportationReview}%
  \BibitemOpen
  \bibfield  {author} {\bibinfo {author} {\bibfnamefont {S.}~\bibnamefont
  {Pirandola}}, \bibinfo {author} {\bibfnamefont {J.}~\bibnamefont {Eisert}},
  \bibinfo {author} {\bibfnamefont {C.}~\bibnamefont {Weedbrook}}, \bibinfo
  {author} {\bibfnamefont {A.}~\bibnamefont {Furusawa}}, \ and\ \bibinfo
  {author} {\bibfnamefont {S.~L.}\ \bibnamefont {Braunstein}},\ }\bibfield
  {title} {\enquote {\bibinfo {title} {Advances in quantum teleportation},}\
  }\href {\doibase 10.1038/nphoton.2015.154} {\bibfield  {journal} {\bibinfo
  {journal} {Nature Phot.}\ }\textbf {\bibinfo {volume} {9}},\ \bibinfo {pages}
  {641--652} (\bibinfo {year} {2015})}\BibitemShut {NoStop}%
\bibitem [{\citenamefont {Portmann}\ and\ \citenamefont
  {Renner}(2021)}]{Renner2021}%
  \BibitemOpen
  \bibfield  {author} {\bibinfo {author} {\bibfnamefont {C.}~\bibnamefont
  {Portmann}}\ and\ \bibinfo {author} {\bibfnamefont {R.}~\bibnamefont
  {Renner}},\ }\bibfield  {title} {\enquote {\bibinfo {title} {Security in
  quantum cryptography},}\ }\href@noop {} {\bibfield  {journal} {\bibinfo
  {journal} {arXiv:2102.00021}\ } (\bibinfo {year} {2021})}\BibitemShut
  {NoStop}%
\bibitem [{\citenamefont {Khatri}\ and\ \citenamefont
  {Wilde}(2021)}]{Principles}%
  \BibitemOpen
  \bibfield  {author} {\bibinfo {author} {\bibfnamefont {S.}~\bibnamefont
  {Khatri}}\ and\ \bibinfo {author} {\bibfnamefont {M.~M.}\ \bibnamefont
  {Wilde}},\ }\bibfield  {title} {\enquote {\bibinfo {title} {Principles of
  quantum communication theory: A modern approach},}\ }\href@noop {} {\bibfield
   {journal} {\bibinfo  {journal} {arXiv:2011.04672}\ } (\bibinfo {year}
  {2021})}\BibitemShut {NoStop}%
\bibitem [{\citenamefont {Werner}(2001)}]{Werner2001}%
  \BibitemOpen
  \bibfield  {author} {\bibinfo {author} {\bibfnamefont {R.~F.}\ \bibnamefont
  {Werner}},\ }\enquote {\bibinfo {title} {Quantum information theory --- an
  invitation},}\ in\ \href {\doibase 10.1007/3-540-44678-8_2} {\emph {\bibinfo
  {booktitle} {Quantum Information: An Introduction to Basic Theoretical
  Concepts and Experiments}}}\ (\bibinfo  {publisher} {Springer Berlin
  Heidelberg},\ \bibinfo {address} {Berlin, Heidelberg},\ \bibinfo {year}
  {2001})\ pp.\ \bibinfo {pages} {14--57}\BibitemShut {NoStop}%
\bibitem [{\citenamefont {Sidhu}\ \emph {et~al.}(2021)\citenamefont {Sidhu},
  \citenamefont {Joshi}, \citenamefont {Gündoğan}, \citenamefont {Brougham},
  \citenamefont {Lowndes}, \citenamefont {Mazzarella}, \citenamefont {Krutzik},
  \citenamefont {Mohapatra}, \citenamefont {Dequal}, \citenamefont {Vallone},
  \citenamefont {Villoresi}, \citenamefont {Ling}, \citenamefont {Jennewein},
  \citenamefont {Mohageg}, \citenamefont {Rarity}, \citenamefont {Fuentes},
  \citenamefont {Pirandola},\ and\ \citenamefont {Oi}}]{RevSpace}%
  \BibitemOpen
  \bibfield  {author} {\bibinfo {author} {\bibfnamefont {J.~S.}\ \bibnamefont
  {Sidhu}}, \bibinfo {author} {\bibfnamefont {S.~K.}\ \bibnamefont {Joshi}},
  \bibinfo {author} {\bibfnamefont {M.}~\bibnamefont {Gündoğan}}, \bibinfo
  {author} {\bibfnamefont {T.}~\bibnamefont {Brougham}}, \bibinfo {author}
  {\bibfnamefont {D.}~\bibnamefont {Lowndes}}, \bibinfo {author} {\bibfnamefont
  {L.}~\bibnamefont {Mazzarella}}, \bibinfo {author} {\bibfnamefont
  {M.}~\bibnamefont {Krutzik}}, \bibinfo {author} {\bibfnamefont
  {S.}~\bibnamefont {Mohapatra}}, \bibinfo {author} {\bibfnamefont
  {D.}~\bibnamefont {Dequal}}, \bibinfo {author} {\bibfnamefont {G.}\
  \bibnamefont {Vallone}}, \bibinfo {author} {\bibfnamefont {P.}~\bibnamefont
  {Villoresi}}, \bibinfo {author} {\bibfnamefont {A.}~\bibnamefont {Ling}},
  \bibinfo {author} {\bibfnamefont {T.}~\bibnamefont {Jennewein}}, \bibinfo
  {author} {\bibfnamefont {M.}~\bibnamefont {Mohageg}}, \bibinfo {author}
  {\bibfnamefont {J.~G.}\ \bibnamefont {Rarity}}, \bibinfo {author}
  {\bibfnamefont {I.}~\bibnamefont {Fuentes}}, \bibinfo {author} {\bibfnamefont
  {S.}~\bibnamefont {Pirandola}}, \ and\ \bibinfo {author} {\bibfnamefont
  {D.~K.~L.}\ \bibnamefont {Oi}},\ }\bibfield  {title} {\enquote {\bibinfo
  {title} {Advances in space quantum communications},}\ }\href {\doibase
  https://doi.org/10.1049/qtc2.12015} {\bibfield  {journal} {\bibinfo
  {journal} {IET Quant. Comm.}\ }\textbf {\bibinfo {volume} {93}},\ \bibinfo
  {pages} {1– 36} (\bibinfo {year} {2021})}\BibitemShut {NoStop}%
\bibitem [{\citenamefont {Razavi}(2018)}]{MohsenBook}%
  \BibitemOpen
  \bibfield  {author} {\bibinfo {author} {\bibfnamefont {M.}~\bibnamefont
  {Razavi}},\ }\href {\doibase 10.1088/978-1-6817-4653-1} {\emph {\bibinfo
  {title} {An Introduction to Quantum Communications Networks}}},\ 2053-2571\
  (\bibinfo  {publisher} {Morgan \& Claypool Publishers},\ \bibinfo {year}
  {2018})\BibitemShut {NoStop}%
\bibitem [{\citenamefont {Pirandola}\ \emph {et~al.}(2017)\citenamefont
  {Pirandola}, \citenamefont {Laurenza}, \citenamefont {Ottaviani},\ and\
  \citenamefont {Banchi}}]{PLOB}%
  \BibitemOpen
  \bibfield  {author} {\bibinfo {author} {\bibfnamefont {S.}~\bibnamefont
  {Pirandola}}, \bibinfo {author} {\bibfnamefont {R.}~\bibnamefont {Laurenza}},
  \bibinfo {author} {\bibfnamefont {C.}~\bibnamefont {Ottaviani}}, \ and\
  \bibinfo {author} {\bibfnamefont {L.}~\bibnamefont {Banchi}},\ }\bibfield
  {title} {\enquote {\bibinfo {title} {{Fundamental limits of repeaterless
  quantum communications}},}\ }\href {\doibase 10.1038/ncomms15043} {\bibfield
  {journal} {\bibinfo  {journal} {Nature Comm.}\ }\textbf {\bibinfo {volume}
  {8}},\ \bibinfo {pages} {15043} (\bibinfo {year} {2017})}\BibitemShut
  {NoStop}%
\bibitem [{\citenamefont {Sangouard}\ \emph {et~al.}(2011)\citenamefont
  {Sangouard}, \citenamefont {Simon}, \citenamefont {de~Riedmatten},\ and\
  \citenamefont {Gisin}}]{RevModPhys.83.33}%
  \BibitemOpen
  \bibfield  {author} {\bibinfo {author} {\bibfnamefont {N.}~\bibnamefont
  {Sangouard}}, \bibinfo {author} {\bibfnamefont {C.}~\bibnamefont {Simon}},
  \bibinfo {author} {\bibfnamefont {H.}~\bibnamefont {de~Riedmatten}}, \ and\
  \bibinfo {author} {\bibfnamefont {N.}~\bibnamefont {Gisin}},\ }\bibfield
  {title} {\enquote {\bibinfo {title} {Quantum repeaters based on atomic
  ensembles and linear optics},}\ }\href {\doibase 10.1103/RevModPhys.83.33}
  {\bibfield  {journal} {\bibinfo  {journal} {Rev. Mod. Phys.}\ }\textbf
  {\bibinfo {volume} {83}},\ \bibinfo {pages} {33--80} (\bibinfo {year}
  {2011})}\BibitemShut {NoStop}%
\bibitem [{\citenamefont {Briegel}\ \emph {et~al.}(1998)\citenamefont
  {Briegel}, \citenamefont {D{\"u}r}, \citenamefont {Cirac},\ and\
  \citenamefont {Zoller}}]{Briegel}%
  \BibitemOpen
  \bibfield  {author} {\bibinfo {author} {\bibfnamefont {H.-J.}\ \bibnamefont
  {Briegel}}, \bibinfo {author} {\bibfnamefont {W.}~\bibnamefont {D{\"u}r}},
  \bibinfo {author} {\bibfnamefont {J.}~\bibnamefont {Cirac}}, \ and\ \bibinfo
  {author} {\bibfnamefont {P.}~\bibnamefont {Zoller}},\ }\bibfield  {title}
  {\enquote {\bibinfo {title} {{Quantum repeaters: The role of imperfect local
  operations in quantum communication}},}\ }\href {\doibase
  10.1103/PhysRevLett.81.5932} {\bibfield  {journal} {\bibinfo  {journal}
  {Phys. Rev. Lett.}\ }\textbf {\bibinfo {volume} {81}},\ \bibinfo {pages}
  {5932--5935} (\bibinfo {year} {1998})}\BibitemShut {NoStop}%
\bibitem [{\citenamefont {Pirandola}(2019)}]{PirNetwork}%
  \BibitemOpen
  \bibfield  {author} {\bibinfo {author} {\bibfnamefont {S.}~\bibnamefont
  {Pirandola}},\ }\bibfield  {title} {\enquote {\bibinfo {title} {{End-to-end
  capacities of a quantum communication network}},}\ }\href {\doibase
  10.1038/s42005-019-0147-3} {\bibfield  {journal} {\bibinfo  {journal}
  {Commun. Phys.}\ }\textbf {\bibinfo {volume} {2}},\ \bibinfo {pages} {51}
  (\bibinfo {year} {2019})},\ \bibinfo {note} {see also
  arXiv:1601.00966}\BibitemShut {NoStop}%
\bibitem [{\citenamefont {Kimble}(2008)}]{QuantumInternet}%
  \BibitemOpen
  \bibfield  {author} {\bibinfo {author} {\bibfnamefont {H.~J.}\ \bibnamefont
  {Kimble}},\ }\bibfield  {title} {\enquote {\bibinfo {title} {The quantum
  internet},}\ }\href {\doibase 10.1038/nature07127} {\bibfield  {journal}
  {\bibinfo  {journal} {Nature}\ }\textbf {\bibinfo {volume} {453}},\ \bibinfo
  {pages} {1023--1030} (\bibinfo {year} {2008})}\BibitemShut {NoStop}%
\bibitem [{\citenamefont {Satoh}\ \emph {et~al.}(2016)\citenamefont {Satoh},
  \citenamefont {Ishizaki}, \citenamefont {Nagayama},\ and\ \citenamefont
  {Van~Meter}}]{PhysRevA.93.032302}%
  \BibitemOpen
  \bibfield  {author} {\bibinfo {author} {\bibfnamefont {T.}~\bibnamefont
  {Satoh}}, \bibinfo {author} {\bibfnamefont {K.}~\bibnamefont {Ishizaki}},
  \bibinfo {author} {\bibfnamefont {S.}~\bibnamefont {Nagayama}}, \ and\
  \bibinfo {author} {\bibfnamefont {R.}~\bibnamefont {Van~Meter}},\ }\bibfield
  {title} {\enquote {\bibinfo {title} {Analysis of quantum network coding for
  realistic repeater networks},}\ }\href {\doibase 10.1103/PhysRevA.93.032302}
  {\bibfield  {journal} {\bibinfo  {journal} {Phys. Rev. A}\ }\textbf {\bibinfo
  {volume} {93}},\ \bibinfo {pages} {032302} (\bibinfo {year}
  {2016})}\BibitemShut {NoStop}%
\bibitem [{\citenamefont {Epping}\ \emph {et~al.}(2016)\citenamefont {Epping},
  \citenamefont {Kampermann},\ and\ \citenamefont {Bru\ss}}]{EppingA}%
  \BibitemOpen
  \bibfield  {author} {\bibinfo {author} {\bibfnamefont {M.}~\bibnamefont
  {Epping}}, \bibinfo {author} {\bibfnamefont {H.}~\bibnamefont {Kampermann}},
  \ and\ \bibinfo {author} {\bibnamefont {Bru\ss}},\ }\bibfield  {title}
  {\enquote {\bibinfo {title} {Large-scale quantum networks based on graphs},}\
  }\href {\doibase 10.1088/1367-2630/18/5/053036} {\bibfield  {journal}
  {\bibinfo  {journal} {New J. Phys.}\ }\textbf {\bibinfo {volume} {18}},\
  \bibinfo {pages} {053036} (\bibinfo {year} {2016})}\BibitemShut {NoStop}%
\bibitem [{\citenamefont {Hahn}\ \emph {et~al.}(2019)\citenamefont {Hahn},
  \citenamefont {Pappa},\ and\ \citenamefont {Eisert}}]{HahnPappaEisert}%
  \BibitemOpen
  \bibfield  {author} {\bibinfo {author} {\bibfnamefont {F.}~\bibnamefont
  {Hahn}}, \bibinfo {author} {\bibfnamefont {A.}~\bibnamefont {Pappa}}, \ and\
  \bibinfo {author} {\bibfnamefont {J.}~\bibnamefont {Eisert}},\ }\bibfield
  {title} {\enquote {\bibinfo {title} {Quantum network routing and local
  complementation},}\ }\href {\doibase 10.1038/s41534-019-0191-6} {\bibfield
  {journal} {\bibinfo  {journal} {npj Quantum Inf.}\ }\textbf {\bibinfo
  {volume} {5}},\ \bibinfo {pages} {76} (\bibinfo {year} {2019})}\BibitemShut
  {NoStop}%
\bibitem [{\citenamefont {Coopmans}\ \emph {et~al.}(2021)\citenamefont
  {Coopmans}, \citenamefont {Knegjens}, \citenamefont {Dahlberg}, \citenamefont
  {Maier}, \citenamefont {Nijsten}, \citenamefont {de~Oliveira~Filho},
  \citenamefont {Papendrecht}, \citenamefont {Rabbie}, \citenamefont
  {Rozpedek}, \citenamefont {Skrzypczyk}, \citenamefont {Wubben}, \citenamefont
  {de~Jong}, \citenamefont {Podareanu}, \citenamefont {Torres-Knoop},
  \citenamefont {Elkouss},\ and\ \citenamefont {Wehner}}]{NetSquid}%
  \BibitemOpen
  \bibfield  {author} {\bibinfo {author} {\bibfnamefont {T.}~\bibnamefont
  {Coopmans}}, \bibinfo {author} {\bibfnamefont {R.}~\bibnamefont {Knegjens}},
  \bibinfo {author} {\bibfnamefont {A.}~\bibnamefont {Dahlberg}}, \bibinfo
  {author} {\bibfnamefont {D.}~\bibnamefont {Maier}}, \bibinfo {author}
  {\bibfnamefont {L.}~\bibnamefont {Nijsten}}, \bibinfo {author} {\bibfnamefont
  {J.}~\bibnamefont {de~Oliveira~Filho}}, \bibinfo {author} {\bibfnamefont
  {M.}~\bibnamefont {Papendrecht}}, \bibinfo {author} {\bibfnamefont
  {J.}~\bibnamefont {Rabbie}}, \bibinfo {author} {\bibfnamefont
  {F.}~\bibnamefont {Rozpedek}}, \bibinfo {author} {\bibfnamefont
  {M.}~\bibnamefont {Skrzypczyk}}, \bibinfo {author} {\bibfnamefont
  {L.}~\bibnamefont {Wubben}}, \bibinfo {author} {\bibfnamefont
  {W.}~\bibnamefont {de~Jong}}, \bibinfo {author} {\bibfnamefont
  {D.}~\bibnamefont {Podareanu}}, \bibinfo {author} {\bibfnamefont
  {A.}~\bibnamefont {Torres-Knoop}}, \bibinfo {author} {\bibfnamefont
  {D.}~\bibnamefont {Elkouss}}, \ and\ \bibinfo {author} {\bibfnamefont
  {S.}~\bibnamefont {Wehner}},\ }\bibfield  {title} {\enquote {\bibinfo {title}
  {{NetSquid, a NETwork Simulator for QUantum Information using Discrete
  events}},}\ }\href {\doibase 10.1038/s42005-021-00647-8} {\bibfield
  {journal} {\bibinfo  {journal} {Commun. Phys.}\ }\textbf {\bibinfo {volume}
  {4}},\ \bibinfo {pages} {164} (\bibinfo {year} {2021})}\BibitemShut {NoStop}%
\bibitem [{\citenamefont {Goodenough}\ \emph {et~al.}(2021)\citenamefont
  {Goodenough}, \citenamefont {Elkouss},\ and\ \citenamefont
  {Wehner}}]{PhysRevA.103.032610}%
  \BibitemOpen
  \bibfield  {author} {\bibinfo {author} {\bibfnamefont {K.}~\bibnamefont
  {Goodenough}}, \bibinfo {author} {\bibfnamefont {D.}~\bibnamefont {Elkouss}},
  \ and\ \bibinfo {author} {\bibfnamefont {S.}~\bibnamefont {Wehner}},\
  }\bibfield  {title} {\enquote {\bibinfo {title} {Optimizing repeater schemes
  for the quantum internet},}\ }\href {\doibase 10.1103/PhysRevA.103.032610}
  {\bibfield  {journal} {\bibinfo  {journal} {Phys. Rev. A}\ }\textbf {\bibinfo
  {volume} {103}},\ \bibinfo {pages} {032610} (\bibinfo {year}
  {2021})}\BibitemShut {NoStop}%
\bibitem [{\citenamefont {Eisert}\ and\ \citenamefont
  {Wolf}(2007)}]{GaussianChannel}%
  \BibitemOpen
  \bibfield  {author} {\bibinfo {author} {\bibfnamefont {J.}~\bibnamefont
  {Eisert}}\ and\ \bibinfo {author} {\bibfnamefont {M.~M.}\ \bibnamefont
  {Wolf}},\ }\enquote {\bibinfo {title} {Gaussian quantum channels},}\ in\
  \href {\doibase 10.1142/p489} {\emph {\bibinfo {booktitle} {Quantum
  Information with Continous Variables of Atoms and Light}}}\ (\bibinfo
  {publisher} {Imperial College Press},\ \bibinfo {address} {London},\ \bibinfo
  {year} {2007})\ pp.\ \bibinfo {pages} {23--42},\ \bibinfo {note}
  {arXiv:quant-ph/0505151}\BibitemShut {NoStop}%
\bibitem [{not()}]{note2}%
  \BibitemOpen
  \href@noop {} {}\bibinfo {note} {The bound $B$ in Eq.~(\ref{chainCap1}) in
  terms of bits per channel use can be translated into a bound in terms of bits
  per second once we consider a clock $C$ for the generation of the states at
  the various nodes. This can be expressed in terms of uses per second. When we
  combine the two parameters as $B\times C$ we then get the throughput in terms
  of bits per second. Note that this reasoning assumes the ideal case of
  identical clocks for all the nodes; in the case where the clocks are
  different, then we can take the maximum value of them since we are computing
  an upper bound.}\BibitemShut {Stop}%
\bibitem [{\citenamefont {Hosseini}\ \emph {et~al.}(2011)\citenamefont
  {Hosseini}, \citenamefont {Sparkes}, \citenamefont {Campbell}, \citenamefont
  {Lam},\ and\ \citenamefont {Buchler}}]{Mem}%
  \BibitemOpen
  \bibfield  {author} {\bibinfo {author} {\bibfnamefont {M.}~\bibnamefont
  {Hosseini}}, \bibinfo {author} {\bibfnamefont {B.~M.}\ \bibnamefont
  {Sparkes}}, \bibinfo {author} {\bibfnamefont {G.}~\bibnamefont {Campbell}},
  \bibinfo {author} {\bibfnamefont {P.~K.}\ \bibnamefont {Lam}}, \ and\
  \bibinfo {author} {\bibfnamefont {B.~C.}\ \bibnamefont {Buchler}},\
  }\bibfield  {title} {\enquote {\bibinfo {title} {{High efficiency coherent
  optical memory with warm rubidium vapour}},}\ }\href {\doibase
  10.1038/ncomms1175} {\bibfield  {journal} {\bibinfo  {journal} {Nature
  Comm.}\ }\textbf {\bibinfo {volume} {2}},\ \bibinfo {pages} {174} (\bibinfo
  {year} {2011})}\BibitemShut {NoStop}%
\bibitem [{\citenamefont {Bernardes}\ \emph {et~al.}(2011)\citenamefont
  {Bernardes}, \citenamefont {Praxmeyer},\ and\ \citenamefont {van
  Loock}}]{Bernardes:2011ij}%
  \BibitemOpen
  \bibfield  {author} {\bibinfo {author} {\bibfnamefont {N.}~\bibnamefont
  {Bernardes}}, \bibinfo {author} {\bibfnamefont {L.}~\bibnamefont
  {Praxmeyer}}, \ and\ \bibinfo {author} {\bibfnamefont {P.}~\bibnamefont {van
  Loock}},\ }\bibfield  {title} {\enquote {\bibinfo {title} {{Rate analysis for
  a hybrid quantum repeater}},}\ }\href {\doibase 10.1103/PhysRevA.83.012323}
  {\bibfield  {journal} {\bibinfo  {journal} {Phys. Rev. A}\ }\textbf {\bibinfo
  {volume} {83}},\ \bibinfo {pages} {012323} (\bibinfo {year}
  {2011})}\BibitemShut {NoStop}%
\bibitem [{\citenamefont {Shchukin}\ \emph {et~al.}(2019)\citenamefont
  {Shchukin}, \citenamefont {Schmidt},\ and\ \citenamefont {van
  Loock}}]{Shchukin:2019gs}%
  \BibitemOpen
  \bibfield  {author} {\bibinfo {author} {\bibfnamefont {E.}~\bibnamefont
  {Shchukin}}, \bibinfo {author} {\bibfnamefont {F.}~\bibnamefont {Schmidt}}, \
  and\ \bibinfo {author} {\bibfnamefont {P.}~\bibnamefont {van Loock}},\
  }\bibfield  {title} {\enquote {\bibinfo {title} {{Waiting time in quantum
  repeaters with probabilistic entanglement swapping}},}\ }\href {\doibase
  10.1103/PhysRevA.100.032322} {\bibfield  {journal} {\bibinfo  {journal}
  {Phys. Rev. A}\ }\textbf {\bibinfo {volume} {100}},\ \bibinfo {pages}
  {032322} (\bibinfo {year} {2019})}\BibitemShut {NoStop}%
\bibitem [{\citenamefont {van Loock}\ \emph {et~al.}(2020)\citenamefont {van
  Loock}, \citenamefont {Alt}, \citenamefont {Becher}, \citenamefont {Benson},
  \citenamefont {Boche}, \citenamefont {Deppe}, \citenamefont {Eschner},
  \citenamefont {H{\"o}fling}, \citenamefont {Meschede}, \citenamefont
  {Michler}, \citenamefont {Schmidt},\ and\ \citenamefont
  {Weinfurter}}]{vanLoock:2020gt}%
  \BibitemOpen
  \bibfield  {author} {\bibinfo {author} {\bibfnamefont {P.}~\bibnamefont {van
  Loock}}, \bibinfo {author} {\bibfnamefont {W.}~\bibnamefont {Alt}}, \bibinfo
  {author} {\bibfnamefont {C.}~\bibnamefont {Becher}}, \bibinfo {author}
  {\bibfnamefont {O.}~\bibnamefont {Benson}}, \bibinfo {author} {\bibfnamefont
  {H.}~\bibnamefont {Boche}}, \bibinfo {author} {\bibfnamefont
  {C.}~\bibnamefont {Deppe}}, \bibinfo {author} {\bibfnamefont
  {J.}~\bibnamefont {Eschner}}, \bibinfo {author} {\bibfnamefont
  {S.}~\bibnamefont {H{\"o}fling}}, \bibinfo {author} {\bibfnamefont
  {D.}~\bibnamefont {Meschede}}, \bibinfo {author} {\bibfnamefont
  {P.}~\bibnamefont {Michler}}, \bibinfo {author} {\bibfnamefont
  {F.}~\bibnamefont {Schmidt}}, \ and\ \bibinfo {author} {\bibfnamefont
  {H.}~\bibnamefont {Weinfurter}},\ }\bibfield  {title} {\enquote {\bibinfo
  {title} {{Extending quantum links: Modules for fiber- and memory-based
  quantum repeaters}},}\ }\href {\doibase 10.48550/arXiv.1912.10123} {\bibfield
   {journal} {\bibinfo  {journal} {Adv. Quant. Tech.}\ }\textbf {\bibinfo
  {volume} {3}},\ \bibinfo {pages} {1900141} (\bibinfo {year}
  {2020})}\BibitemShut {NoStop}%
\bibitem [{\citenamefont {Bhaskar}\ \emph {et~al.}(2020)\citenamefont
  {Bhaskar}, \citenamefont {Riedinger}, \citenamefont {Machielse},
  \citenamefont {Levonian}, \citenamefont {Nguyen}, \citenamefont {Knall},
  \citenamefont {Park}, \citenamefont {Englund}, \citenamefont {Lon{\v c}ar},
  \citenamefont {Sukachev},\ and\ \citenamefont {Lukin}}]{Bhaskar:2020gh}%
  \BibitemOpen
  \bibfield  {author} {\bibinfo {author} {\bibfnamefont {M.~K.}\ \bibnamefont
  {Bhaskar}}, \bibinfo {author} {\bibfnamefont {R.}~\bibnamefont {Riedinger}},
  \bibinfo {author} {\bibfnamefont {B.}~\bibnamefont {Machielse}}, \bibinfo
  {author} {\bibfnamefont {D.~S.}\ \bibnamefont {Levonian}}, \bibinfo {author}
  {\bibfnamefont {C.~T.}\ \bibnamefont {Nguyen}}, \bibinfo {author}
  {\bibfnamefont {E.~N.}\ \bibnamefont {Knall}}, \bibinfo {author}
  {\bibfnamefont {H.}~\bibnamefont {Park}}, \bibinfo {author} {\bibfnamefont
  {D.}~\bibnamefont {Englund}}, \bibinfo {author} {\bibfnamefont
  {M.}~\bibnamefont {Lon{\v c}ar}}, \bibinfo {author} {\bibfnamefont {D.~D.}\
  \bibnamefont {Sukachev}}, \ and\ \bibinfo {author} {\bibfnamefont {M.~D.}\
  \bibnamefont {Lukin}},\ }\bibfield  {title} {\enquote {\bibinfo {title}
  {{Experimental demonstration of memory-enhanced quantum communication}},}\
  }\href {\doibase 10.1038/s41586-020-2103-5} {\bibfield  {journal} {\bibinfo
  {journal} {Nature}\ }\textbf {\bibinfo {volume} {580}},\ \bibinfo {pages}
  {60--64} (\bibinfo {year} {2020})}\BibitemShut {NoStop}%
\bibitem [{\citenamefont {Pu}\ \emph {et~al.}(2021)\citenamefont {Pu},
  \citenamefont {Zhang}, \citenamefont {Wu}, \citenamefont {Jiang},
  \citenamefont {Chang}, \citenamefont {Li},\ and\ \citenamefont
  {Duan}}]{Pu:2021tr}%
  \BibitemOpen
  \bibfield  {author} {\bibinfo {author} {\bibfnamefont {Y.}~\bibnamefont
  {Pu}}, \bibinfo {author} {\bibfnamefont {S.}~\bibnamefont {Zhang}}, \bibinfo
  {author} {\bibfnamefont {Y.}~\bibnamefont {Wu}}, \bibinfo {author}
  {\bibfnamefont {N.}~\bibnamefont {Jiang}}, \bibinfo {author} {\bibfnamefont
  {W.}~\bibnamefont {Chang}}, \bibinfo {author} {\bibfnamefont
  {C.}~\bibnamefont {Li}}, \ and\ \bibinfo {author} {\bibfnamefont
  {L.}~\bibnamefont {Duan}},\ }\bibfield  {title} {\enquote {\bibinfo {title}
  {{Experimental demonstration of memory-enhanced scaling for entanglement
  connection of quantum repeater segments}},}\ }\href {\doibase
  10.1038/s41566-021-00764-4} {\bibfield  {journal} {\bibinfo  {journal}
  {Nature Phot.}\ }\textbf {\bibinfo {volume} {15}},\ \bibinfo {pages}
  {374--378} (\bibinfo {year} {2021})}\BibitemShut {NoStop}%
\bibitem [{\citenamefont {Campbell}\ and\ \citenamefont
  {Eisert}(2012)}]{Campbell:2012bq}%
  \BibitemOpen
  \bibfield  {author} {\bibinfo {author} {\bibfnamefont {E.}~\bibnamefont
  {Campbell}}\ and\ \bibinfo {author} {\bibfnamefont {J.}~\bibnamefont
  {Eisert}},\ }\bibfield  {title} {\enquote {\bibinfo {title} {{Gaussification
  and entanglement distillation of continuous-variable systems: A unifying
  picture}},}\ }\href {\doibase /10.1103/PhysRevLett.108.020501} {\bibfield
  {journal} {\bibinfo  {journal} {Phys. Rev. Lett.}\ }\textbf {\bibinfo
  {volume} {108}},\ \bibinfo {pages} {020501} (\bibinfo {year}
  {2012})}\BibitemShut {NoStop}%
\bibitem [{\citenamefont {Dias}\ and\ \citenamefont
  {Ralph}(2017)}]{Dias:2017jk}%
  \BibitemOpen
  \bibfield  {author} {\bibinfo {author} {\bibfnamefont {J.}~\bibnamefont
  {Dias}}\ and\ \bibinfo {author} {\bibfnamefont {T.~C.}\ \bibnamefont
  {Ralph}},\ }\bibfield  {title} {\enquote {\bibinfo {title} {{Quantum
  repeaters using continuous-variable teleportation}},}\ }\href {\doibase
  10.1103/PhysRevA.95.022312} {\bibfield  {journal} {\bibinfo  {journal} {Phys.
  Rev. A}\ }\textbf {\bibinfo {volume} {95}},\ \bibinfo {pages} {022312}
  (\bibinfo {year} {2017})}\BibitemShut {NoStop}%
\bibitem [{\citenamefont {Furrer}\ and\ \citenamefont
  {Munro}(2018)}]{Furrer:2018im}%
  \BibitemOpen
  \bibfield  {author} {\bibinfo {author} {\bibfnamefont {F.}~\bibnamefont
  {Furrer}}\ and\ \bibinfo {author} {\bibfnamefont {W.~J.}\ \bibnamefont
  {Munro}},\ }\bibfield  {title} {\enquote {\bibinfo {title} {{Repeaters for
  continuous-variable quantum communication}},}\ }\href {\doibase
  10.1103/PhysRevA.98.032335} {\bibfield  {journal} {\bibinfo  {journal} {Phys.
  Rev. A}\ }\textbf {\bibinfo {volume} {98}},\ \bibinfo {pages} {032335}
  (\bibinfo {year} {2018})}\BibitemShut {NoStop}%
\bibitem [{\citenamefont {Dias}\ \emph {et~al.}(2019)\citenamefont {Dias},
  \citenamefont {Munro}, \citenamefont {Ralph},\ and\ \citenamefont
  {Nemoto}}]{Dias:2019tz}%
  \BibitemOpen
  \bibfield  {author} {\bibinfo {author} {\bibfnamefont {J.}~\bibnamefont
  {Dias}}, \bibinfo {author} {\bibfnamefont {W.~J.}\ \bibnamefont {Munro}},
  \bibinfo {author} {\bibfnamefont {T.~C.}\ \bibnamefont {Ralph}}, \ and\
  \bibinfo {author} {\bibfnamefont {K.}~\bibnamefont {Nemoto}},\ }\bibfield
  {title} {\enquote {\bibinfo {title} {{Comparison of entanglement generation
  rates between continuous and discrete variable repeaters}},}\ }\href@noop {}
  {\  (\bibinfo {year} {2019})},\ \Eprint
  {http://arxiv.org/abs/arXiv:1906.06019} {arXiv:1906.06019} \BibitemShut
  {NoStop}%
\bibitem [{\citenamefont {Ralph}\ and\ \citenamefont
  {Lund}(2009)}]{proc-disc-2009}%
  \BibitemOpen
  \bibfield  {author} {\bibinfo {author} {\bibfnamefont {T.~C.}\ \bibnamefont
  {Ralph}}\ and\ \bibinfo {author} {\bibfnamefont {A.~P.}\ \bibnamefont
  {Lund}},\ }\href {\doibase 10.1063/1.3131295} {\bibfield  {journal} {\bibinfo
   {journal} {{\it Quantum Communication Measurement and Computing Proceedings
  of 9th International Conference}}\ ,\ \bibinfo {pages} {155}} (\bibinfo
  {year} {2009})}\BibitemShut {NoStop}%
\bibitem [{\citenamefont {Xiang}\ \emph {et~al.}(2010)\citenamefont {Xiang},
  \citenamefont {Ralph}, \citenamefont {Lund}, \citenamefont {Walk},\ and\
  \citenamefont {Pryde}}]{Xiang:2010ua}%
  \BibitemOpen
  \bibfield  {author} {\bibinfo {author} {\bibfnamefont {G.~Y.}\ \bibnamefont
  {Xiang}}, \bibinfo {author} {\bibfnamefont {T.~C.}\ \bibnamefont {Ralph}},
  \bibinfo {author} {\bibfnamefont {A.~P.}\ \bibnamefont {Lund}}, \bibinfo
  {author} {\bibfnamefont {N.}~\bibnamefont {Walk}}, \ and\ \bibinfo {author}
  {\bibfnamefont {G.~J.}\ \bibnamefont {Pryde}},\ }\bibfield  {title} {\enquote
  {\bibinfo {title} {{Heralded noiseless linear amplification and distillation
  of entanglement}},}\ }\href {\doibase 10.1038/nphoton.2010.35} {\bibfield
  {journal} {\bibinfo  {journal} {Nature Phot.}\ }\textbf {\bibinfo {volume}
  {4}},\ \bibinfo {pages} {316--319} (\bibinfo {year} {2010})}\BibitemShut
  {NoStop}%
\bibitem [{\citenamefont {Pandey}\ \emph {et~al.}(2013)\citenamefont {Pandey},
  \citenamefont {Jiang}, \citenamefont {Combes},\ and\ \citenamefont
  {Caves}}]{Pandey:2013wb}%
  \BibitemOpen
  \bibfield  {author} {\bibinfo {author} {\bibfnamefont {S.}~\bibnamefont
  {Pandey}}, \bibinfo {author} {\bibfnamefont {Z.}~\bibnamefont {Jiang}},
  \bibinfo {author} {\bibfnamefont {J.}~\bibnamefont {Combes}}, \ and\ \bibinfo
  {author} {\bibfnamefont {C.~M.}\ \bibnamefont {Caves}},\ }\bibfield  {title}
  {\enquote {\bibinfo {title} {{Quantum limits on probabilistic amplifiers}},}\
  }\href {\doibase 10.1103/PhysRevA.88.033852} {\bibfield  {journal} {\bibinfo
  {journal} {Phys. Rev. A}\ }\textbf {\bibinfo {volume} {88}},\ \bibinfo
  {pages} {033852} (\bibinfo {year} {2013})}\BibitemShut {NoStop}%
\bibitem [{\citenamefont {Schumacher}\ and\ \citenamefont
  {Nielsen}(1996)}]{Coh1}%
  \BibitemOpen
  \bibfield  {author} {\bibinfo {author} {\bibfnamefont {B.}~\bibnamefont
  {Schumacher}}\ and\ \bibinfo {author} {\bibfnamefont {M.~A.}\ \bibnamefont
  {Nielsen}},\ }\bibfield  {title} {\enquote {\bibinfo {title} {Quantum data
  processing and error correction},}\ }\href {\doibase
  10.1103/PhysRevA.54.2629} {\bibfield  {journal} {\bibinfo  {journal} {Phys.
  Rev. A}\ }\textbf {\bibinfo {volume} {54}},\ \bibinfo {pages} {2629--2635}
  (\bibinfo {year} {1996})}\BibitemShut {NoStop}%
\bibitem [{\citenamefont {Lloyd}(1997)}]{Coh2}%
  \BibitemOpen
  \bibfield  {author} {\bibinfo {author} {\bibfnamefont {S.}~\bibnamefont
  {Lloyd}},\ }\bibfield  {title} {\enquote {\bibinfo {title} {Capacity of the
  noisy quantum channel},}\ }\href {\doibase 10.1103/PhysRevA.55.1613}
  {\bibfield  {journal} {\bibinfo  {journal} {Phys. Rev. A}\ }\textbf {\bibinfo
  {volume} {55}},\ \bibinfo {pages} {1613--1622} (\bibinfo {year}
  {1997})}\BibitemShut {NoStop}%
\bibitem [{\citenamefont {Garc\'{\i}a-Patr\'on}\ \emph
  {et~al.}(2009)\citenamefont {Garc\'{\i}a-Patr\'on}, \citenamefont
  {Pirandola}, \citenamefont {Lloyd},\ and\ \citenamefont {Shapiro}}]{RevCoh1}%
  \BibitemOpen
  \bibfield  {author} {\bibinfo {author} {\bibfnamefont {R.}~\bibnamefont
  {Garc\'{\i}a-Patr\'on}}, \bibinfo {author} {\bibfnamefont {S.}~\bibnamefont
  {Pirandola}}, \bibinfo {author} {\bibfnamefont {S.}~\bibnamefont {Lloyd}}, \
  and\ \bibinfo {author} {\bibfnamefont {J.~H.}\ \bibnamefont {Shapiro}},\
  }\bibfield  {title} {\enquote {\bibinfo {title} {Reverse coherent
  information},}\ }\href {\doibase 10.1103/PhysRevLett.102.210501} {\bibfield
  {journal} {\bibinfo  {journal} {Phys. Rev. Lett.}\ }\textbf {\bibinfo
  {volume} {102}},\ \bibinfo {pages} {210501} (\bibinfo {year}
  {2009})}\BibitemShut {NoStop}%
\bibitem [{\citenamefont {Pirandola}\ \emph {et~al.}(2009)\citenamefont
  {Pirandola}, \citenamefont {Garc\'{\i}a-Patr\'on}, \citenamefont
  {Braunstein},\ and\ \citenamefont {Lloyd}}]{RevCoh2}%
  \BibitemOpen
  \bibfield  {author} {\bibinfo {author} {\bibfnamefont {S.}~\bibnamefont
  {Pirandola}}, \bibinfo {author} {\bibfnamefont {R.}~\bibnamefont
  {Garc\'{\i}a-Patr\'on}}, \bibinfo {author} {\bibfnamefont {S.~L.}\
  \bibnamefont {Braunstein}}, \ and\ \bibinfo {author} {\bibfnamefont
  {S.}~\bibnamefont {Lloyd}},\ }\bibfield  {title} {\enquote {\bibinfo {title}
  {Direct and reverse secret-key capacities of a quantum channel},}\ }\href
  {\doibase 10.1103/PhysRevLett.102.050503} {\bibfield  {journal} {\bibinfo
  {journal} {Phys. Rev. Lett.}\ }\textbf {\bibinfo {volume} {102}},\ \bibinfo
  {pages} {050503} (\bibinfo {year} {2009})}\BibitemShut {NoStop}%
\bibitem [{\citenamefont {Devetak}\ \emph {et~al.}(2006)\citenamefont
  {Devetak}, \citenamefont {Junge}, \citenamefont {King},\ and\ \citenamefont
  {Ruskai}}]{devetakcb}%
  \BibitemOpen
  \bibfield  {author} {\bibinfo {author} {\bibfnamefont {I.}~\bibnamefont
  {Devetak}}, \bibinfo {author} {\bibfnamefont {M.}~\bibnamefont {Junge}},
  \bibinfo {author} {\bibfnamefont {C.}~\bibnamefont {King}}, \ and\ \bibinfo
  {author} {\bibfnamefont {M.~B.}\ \bibnamefont {Ruskai}},\ }\bibfield  {title}
  {\enquote {\bibinfo {title} {{Multiplicativity of completely bounded p-norms
  implies a new additivity result}},}\ }\href {\doibase
  10.1007/s00220-006-0034-0} {\bibfield  {journal} {\bibinfo  {journal}
  {Commun. Math. Phys.}\ }\textbf {\bibinfo {volume} {266}},\ \bibinfo {pages}
  {37--63} (\bibinfo {year} {2006})}\BibitemShut {NoStop}%
\bibitem [{\citenamefont {Hayashi}(2017)}]{hayashipseudo}%
  \BibitemOpen
  \bibfield  {author} {\bibinfo {author} {\bibfnamefont {M.}~\bibnamefont
  {Hayashi}},\ }\bibfield  {title} {\enquote {\bibinfo {title} {{Quantum
  information theory: Mathematical foundation}},}\ }\href {\doibase
  10.1007/978-3-662-49725-8} {\bibfield  {journal} {\bibinfo  {journal}
  {Springer-Verlag}\ } (\bibinfo {year} {2017}),\
  10.1007/978-3-662-49725-8}\BibitemShut {NoStop}%
\bibitem [{\citenamefont {Devetak}\ and\ \citenamefont
  {Winter}(2003)}]{DevWint}%
  \BibitemOpen
  \bibfield  {author} {\bibinfo {author} {\bibfnamefont {I.}~\bibnamefont
  {Devetak}}\ and\ \bibinfo {author} {\bibfnamefont {A.}~\bibnamefont
  {Winter}},\ }\bibfield  {title} {\enquote {\bibinfo {title} {Distillation of
  secret key and entanglement from quantum states},}\ }\href {\doibase
  10.1098/rspa.2004.1372} {\bibfield  {journal} {\bibinfo  {journal} {Proc.
  Roy. Soc. A}\ }\textbf {\bibinfo {volume} {461}},\ \bibinfo {pages} {207}
  (\bibinfo {year} {2003})}\BibitemShut {NoStop}%
\bibitem [{\citenamefont {Vedral}\ and\ \citenamefont {Plenio}(1998)}]{RE2}%
  \BibitemOpen
  \bibfield  {author} {\bibinfo {author} {\bibfnamefont {V.}~\bibnamefont
  {Vedral}}\ and\ \bibinfo {author} {\bibfnamefont {M.~B.}\ \bibnamefont
  {Plenio}},\ }\bibfield  {title} {\enquote {\bibinfo {title} {Entanglement
  measures and purification procedures},}\ }\href {\doibase
  10.1103/PhysRevA.57.1619} {\bibfield  {journal} {\bibinfo  {journal} {Phys.
  Rev. A}\ }\textbf {\bibinfo {volume} {57}},\ \bibinfo {pages} {1619--1633}
  (\bibinfo {year} {1998})}\BibitemShut {NoStop}%
\bibitem [{\citenamefont {Vedral}\ \emph {et~al.}(1997)\citenamefont {Vedral},
  \citenamefont {Plenio}, \citenamefont {Rippin},\ and\ \citenamefont
  {Knight}}]{RE1}%
  \BibitemOpen
  \bibfield  {author} {\bibinfo {author} {\bibfnamefont {V.}~\bibnamefont
  {Vedral}}, \bibinfo {author} {\bibfnamefont {M.~B.}\ \bibnamefont {Plenio}},
  \bibinfo {author} {\bibfnamefont {M.~A.}\ \bibnamefont {Rippin}}, \ and\
  \bibinfo {author} {\bibfnamefont {P.~L.}\ \bibnamefont {Knight}},\ }\bibfield
   {title} {\enquote {\bibinfo {title} {Quantifying entanglement},}\ }\href
  {\doibase 10.1103/PhysRevLett.78.2275} {\bibfield  {journal} {\bibinfo
  {journal} {Phys. Rev. Lett.}\ }\textbf {\bibinfo {volume} {78}},\ \bibinfo
  {pages} {2275--2279} (\bibinfo {year} {1997})}\BibitemShut {NoStop}%
\bibitem [{\citenamefont {Braunstein}\ and\ \citenamefont
  {Kimble}(1998)}]{TeleBK}%
  \BibitemOpen
  \bibfield  {author} {\bibinfo {author} {\bibfnamefont {S.~L.}\ \bibnamefont
  {Braunstein}}\ and\ \bibinfo {author} {\bibfnamefont {H.~J.}\ \bibnamefont
  {Kimble}},\ }\bibfield  {title} {\enquote {\bibinfo {title} {Teleportation of
  continuous quantum variables},}\ }\href {\doibase 10.1103/PhysRevLett.80.869}
  {\bibfield  {journal} {\bibinfo  {journal} {Phys. Rev. Lett.}\ }\textbf
  {\bibinfo {volume} {80}},\ \bibinfo {pages} {869--872} (\bibinfo {year}
  {1998})}\BibitemShut {NoStop}%
\bibitem [{\citenamefont {Braunstein}\ and\ \citenamefont {van
  Loock}(2005)}]{PeterRMP}%
  \BibitemOpen
  \bibfield  {author} {\bibinfo {author} {\bibfnamefont {S.~L.}\ \bibnamefont
  {Braunstein}}\ and\ \bibinfo {author} {\bibfnamefont {P.}~\bibnamefont {van
  Loock}},\ }\bibfield  {title} {\enquote {\bibinfo {title} {Quantum
  information with continuous variables},}\ }\href {\doibase
  10.1103/RevModPhys.77.513} {\bibfield  {journal} {\bibinfo  {journal} {Rev.
  Mod. Phys.}\ }\textbf {\bibinfo {volume} {77}},\ \bibinfo {pages} {513--577}
  (\bibinfo {year} {2005})}\BibitemShut {NoStop}%
\bibitem [{\citenamefont {Bennett}\ \emph {et~al.}(1993)\citenamefont
  {Bennett}, \citenamefont {Brassard}, \citenamefont {Cr\'epeau}, \citenamefont
  {Jozsa}, \citenamefont {Peres},\ and\ \citenamefont
  {Wootters}}]{BennettTELE}%
  \BibitemOpen
  \bibfield  {author} {\bibinfo {author} {\bibfnamefont {C.~H.}\ \bibnamefont
  {Bennett}}, \bibinfo {author} {\bibfnamefont {G.}~\bibnamefont {Brassard}},
  \bibinfo {author} {\bibfnamefont {C.}~\bibnamefont {Cr\'epeau}}, \bibinfo
  {author} {\bibfnamefont {R.}~\bibnamefont {Jozsa}}, \bibinfo {author}
  {\bibfnamefont {A.}~\bibnamefont {Peres}}, \ and\ \bibinfo {author}
  {\bibfnamefont {W.~K.}\ \bibnamefont {Wootters}},\ }\bibfield  {title}
  {\enquote {\bibinfo {title} {{Teleporting an unknown quantum state via dual
  classical and Einstein-Podolsky-Rosen channels}},}\ }\href {\doibase
  10.1103/PhysRevLett.70.1895} {\bibfield  {journal} {\bibinfo  {journal}
  {Phys. Rev. Lett.}\ }\textbf {\bibinfo {volume} {70}},\ \bibinfo {pages}
  {1895--1899} (\bibinfo {year} {1993})}\BibitemShut {NoStop}%
\bibitem [{\citenamefont {Bennett}\ \emph {et~al.}(1996)\citenamefont
  {Bennett}, \citenamefont {DiVincenzo}, \citenamefont {Smolin},\ and\
  \citenamefont {Wootters}}]{BDSW96}%
  \BibitemOpen
  \bibfield  {author} {\bibinfo {author} {\bibfnamefont {C.~H.}\ \bibnamefont
  {Bennett}}, \bibinfo {author} {\bibfnamefont {D.~P.}\ \bibnamefont
  {DiVincenzo}}, \bibinfo {author} {\bibfnamefont {J.~A.}\ \bibnamefont
  {Smolin}}, \ and\ \bibinfo {author} {\bibfnamefont {W.~K.}\ \bibnamefont
  {Wootters}},\ }\bibfield  {title} {\enquote {\bibinfo {title} {{Mixed-state
  entanglement and quantum error correction}},}\ }\href {\doibase
  10.1103/PhysRevA.54.3824} {\bibfield  {journal} {\bibinfo  {journal} {Phys.
  Rev. A}\ }\textbf {\bibinfo {volume} {54}},\ \bibinfo {pages} {3824--3851}
  (\bibinfo {year} {1996})}\BibitemShut {NoStop}%
\end{thebibliography}

%

\end{document}